\documentclass[aip, 10pt,prl,twocolumn,superscriptaddress,showpacs,floatfix]{revtex4}

\pdfoutput=1

\usepackage{graphicx}
\usepackage{graphics}
\usepackage{dcolumn}
\usepackage{mathrsfs,amsmath,bbding,amsfonts,graphics,xcolor}

\usepackage{varioref, subfigure}
\labelformat{equation}{\textup{(#1)}}
\labelformat{figure}{\textup{#1}}
\definecolor{darkblue}{rgb}{0,0.1,0.32}
\usepackage[colorlinks,citecolor=red,urlcolor=green, linkcolor=darkblue, menucolor=blue, breaklinks=true]{hyperref}
\let\orgautoref\autoref

\providecommand{\Autoref}
        {\def\equationautorefname{Equation}%
         \def\figureautorefname{Figure}%
         \def\subfigureautorefname{Figure}%
         \def\Itemautorefname{Item}%
         \def\tableautorefname{Table}%
         \def\chapterautorefname{Chapter}%
         \def\sectionautorefname{Section}%
         \def\subsectionautorefname{Section}%
         \def\subsubsectionautorefname{Section}%
         \def\partautorefname{Part}%
         \orgautoref}

\renewcommand{\autoref}
        {\def\equationautorefname{\!\!}%
         \def\figureautorefname{Fig.}%
         \def\subfigureautorefname{Fig.}%
         \def\Itemautorefname{item}%
         \def\tableautorefname{Table}%
         \def\chapterautorefname{chapter}%
         \def\sectionautorefname{section}%
         \def\subsectionautorefname{section}%
         \def\subsubsectionautorefname{section}%
         \def\partautorefname{part}%
         \orgautoref}

\newcommand{\hm}{\ensuremath{\mathcal{H}\textnormal{-matrix }}}
\newcommand{\hmp}{\ensuremath{\mathcal{H}\textnormal{-matrices }}}
\newcommand{\hms}{\ensuremath{\mathcal{H}^2\textnormal{-matrix }}}

\DeclareMathAlphabet{\mathpzc}{OT1}{pzc}{m}{it}

\usepackage{wasysym}
\usepackage{mathrsfs}

\DeclareMathAlphabet{\mathpzc}{OT1}{pzc}{m}{it}

\newcommand{\norm}[1]{\ensuremath{{|\!| #1 |\!|}}}

\begin{document}

\title{Parametric Hierarchical Matrix Approach for the Wideband Optical Response of Large-Scale Molecular Aggregates}

\author{Davood Ansari-Oghol-Beig}
\email[]{d.ansariogholbeig@neu.edu}
\affiliation{CEM and Photonics Lab, Electrical and Computer Engineering Department, Northeastern University, Boston, MA}

\author{Masoud Rostami}
\affiliation{CEM and Photonics Lab, Electrical and Computer Engineering Department, Northeastern University, Boston, MA}

\author{Ekaterina Chernobrovkina}
\affiliation{CEM and Photonics Lab, Electrical and Computer Engineering Department, Northeastern University, Boston, MA}

\author{Semion K. Saikin}
\affiliation{Department of Chemistry and Chemical Biology, Harvard University, Cambridge, MA}
\affiliation{Department of Physics, Kazan Federal University, 18 Kremlyovskaya Street, Kazan 420008, Russian Federation}

\author{St\'ephanie Valleau}
\affiliation{Department of Chemistry and Chemical Biology, Harvard University, Cambridge, MA}

\author{Hossein Mosallaei}
\email[]{hosseinm@ece.neu.edu}
\affiliation{CEM and Photonics Lab, Electrical and Computer Engineering Department, Northeastern University, Boston, MA}

\author{Al\'an Aspuru-Guzik}
\email[]{alan@aspuru.com}
\affiliation{Department of Chemistry and Chemical Biology, Harvard University, Cambridge, MA}

\date{\today}

\begin{abstract}
Fast and efficient calculations of optical responses using electromagnetic models require computational acceleration and compression techniques. A hierarchical matrix approach is adopted for this purpose. In order to model large-scale molecular structures these methods should be applied over wide frequency spectra. Here we introduce a novel parametric hierarchical matrix method that allows one for a rapid construction of a wideband system representation and enables an efficient wideband solution. We apply the developed method to the modeling of the optical response of bacteriochorophyll tubular aggregates as found in green photosynthetic bacteria. We show that the parametric method can provide one with the frequency and time-domain solutions for structures of the size of $100,000$ molecules, which is comparable to the size of the whole antenna complex in a bacterium. 
The absorption spectrum is calculated and the significance of electrodynamic retardation effects for relatively large structures, i.e. with 
respect to the wavelength of light, is briefly studied.
\end{abstract}


\maketitle 

\section{Introduction}

The prediction of optical properties is one of the main challenges for the theoretical characterization of molecular aggregates \cite{Saikin_Nanoph2013,Blankenship_book,Skolnick_PRL99,Naomi_NanoLett08}. 
The complication originates in the disorder and structural variations that span over a broad length scale and include fluctuations of monomer transition frequencies, domain formation and 
variations in the aggregate shape on the submicron scale \cite{Wurthner2011,Bradley2005,KiDa06_20363_}. The periodic lattice approximation is hardly applicable in this case and one may need to model the complete structure. 
Quantum mechanical methods, for example, open quantum system approaches \cite{RoStEi10_5060_} or quantum mechanics/molecular mechanics methods \cite{ShReVa12_649_}, 
that became popular recently, can characterize aggregate-light interaction in great details. However, the application of these methods to large systems is constrained by the exponential complexity growth with respect 
to the number of monomers composing the structure.

Aggregates of pigments molecules and fluorescent dyes possess distinct optical properties such as strong absorbance and fluorescence, coherent interaction with photons,
and also fast and long-range diffusion of the absorbed energy among the molecules composing the aggregate \cite{Saikin_Nanoph2013}. There are a number of examples of molecular aggregates.
For instance, light-absorbing complexes in plants and photosynthetic bacteria contain aggregates of pigment molecules, chlorophylls and bacteriochlorophylls respectively \cite{Blankenship_book}.
Those structures, constructed by nature, collect and process solar energy with high efficiency. Molecular aggregates can also be grown using self-assembly methods in different shapes including
pseudo one-dimensional chains \cite{Wurthner2011} two-dimensional films \cite{Bradley2005} and nanoscale tubes \cite{KiDa06_20363_, Eisele}. Molecular aggregates can be combined with other photonic structures
such as optical cavities \cite{Skolnick_PRL99} or plasmonic nanoparticles \cite{Naomi_NanoLett08}. Thus, the interest to molecular aggregates as possible light-processing elements grows continuously.

In this context, the classical electrodynamics approach to molecular aggregates \cite{DeVoe_JCP1965,Keller_JCP1986, DDA1}, where molecular excitations are considered as \emph{Hertzian} 
dipoles and quantum properties are described by dipole polarizabilities, can be more convenient in terms of computational effort. This approach is similar to the quantum mechanical \emph{Green}'s function method, where only vibrations of the ground electronic state are taken into account \cite{Roden2009}. 
This approximation, can be understood as a \emph{Galerkin} integral equation (IE) method, in which the trial and test functions are substituted with
\emph{Dirac} delta distributions. 
Besides the polynomial scaling of complexity, the approach opens the venue for the implementation of more efficient acceleration techniques such as the fast multipole method (FMM) \cite{FMM-pedestrain}, the integral equation 
fast \emph{Fourier} transform (IE-FFT)\cite{IE-FFT} and the hierarchical matrix (\hm) method \cite{HMatrix1, HMatrix2}, all widely employed in computational electromagnetics (CEM).

In this study, we introduce a characterization method -- a parametric \hm method -- to model the multi-frequency electromagnetic response of large scale molecular aggregates. 
The method is based on the \hm approach and is applicable to a variety of structures where sweeping over a frequency spectrum is required. As an example, we apply it to calculate stationary and 
time dependent optical responses of tubular aggregates (rolls) of bacteriochlorophylls (BChls). These are the building blocks of the light-absorbing antenna complex -- the chlorosome -- in green 
photosynthetic bacteria. These rolls can consist of tens of thousands of pigments, and combined together the total number of pigments in the chlorosome may exceed hundred of thousands. 
Electronic excitations in these structures propagate on the femtosecond timescale \cite{Taka2012}. Thus, modeling the time dependent response of BChl rolls would require hundreds of frequency domain solutions which clearly 
cannot be attempted without efficient tools for frequency 
sweeping. To demonstrate the efficiency of the parametric \hm method, the time domain (transient) responses of these tubular aggregates are obtained 
via application of fast \emph{Fourier} transform (FFT) to a wideband collection of frequency domain solutions.

A direct implementation of DDA\cite{Yurkin-DDA} results in a system of linear equations with $N$ unknowns. The solution of this system with factorization techniques results in $\mathcal{O}(N^2)$ memory complexity 
and $\mathcal{O}(N^3)$ operation complexity if direct factorization methods such LU-factorization are used. 
Alternatively, if an iterative matrix solution method is utilized, the operation complexity is of order $\mathcal{O}(k N^2)$, where $k$ is the number of required iterations and $\mathcal{O}(N^2)$ is due to the 
complexity of a single matrix-vector multiplication\cite{YousefSaad1}. 
In this respect, all acceleration techniques introduced in computational electromagnetics, provide efficient means to reduce the complexity associated with the latter operation. 
FMM\cite{FMM-pedestrain} and FFT-based accelerators such as IE-FFT\cite{IE-FFT} and are typical examples of such acceleration techniques.

More recently, \hm techniques, i.e. the \hm and the \hms method, were introduced to reduce the computational complexity associated with matrix-vector products 
resulting from the discretization of elliptic differential equations\cite{HMatrix1, HMatrix2}. 
It has been shown that in this case the complexity scales as $\mathcal{O}(N log(N))$ with the size of the matrix. \emph{Banjai} and \emph{Hackbusch} \cite{HMatrixHelmholtz1}, proposed that specifically tailored version of the \hms method can also recover the desirable $\mathcal{O}(N log(N))$ complexity for the discretization of hyperbolic operators that appear, 
for instance, in the solution of electrodynamic (\emph{Maxwell}) equations. 
Moreover, \hmp can be used for effective acceleration of method-of-moments (MoM) solvers involving hyperbolic operators, provided that the problem dimensions are not very large compared 
to the wavelength of the operation\cite{HMatrixHelmholtz1}. The latter condition is usually satisfied for molecular aggregates, as these structures 
are typically smaller or comparable to optical wavelengths.

The rest of the manuscript is organized as follows. 
In \autoref{DDAFrom}, we review the derivation of DDA and its application to optical response of molecular structures.
\Autoref{introduce-HM} gives a quick introduction to hierarchical matrices.
In \autoref{OptimalRep}, we introduce the parametric \hm representation intended for the efficient treatment of multi-frequency DDA problems. 
We discuss errors that appear in parametric \hm representation in \autoref{ErrCon}. In \autoref{Model},
the developed method is used to model optical response of BChl tubular aggregates. Finally, we conclude our study in \autoref{Conc}.

\section{Semiclassical DDA }\label{DDAFrom}

In the following, arbitrary vectors in $\mathbb{R}^3$ are explicitly denoted by the $\vec{\bullet}$ sign and unit vectors in $\mathbb{R}^3$ are denoted by the $\hat{\bullet}$ sign. 
Moreover, the length of the displacement vector $\vec{r}$ and the polarization vector $\vec{p}$ is respectively denoted by $r$ and $p$, dropping the vector sign $\vec{\bullet}$. 
Also, 3-dimensional tensors, i.e. linear transformations $\mathbb{R}^3 \rightarrow \mathbb{R}^3$, are denoted by the $\bar{\bar{\bullet}}$ sign.
According to \emph{Maxwell's} equations the electric field at position $\vec{r}$ can be written as
\begin{equation}
\vec{E}(\vec{r})=-\eta_{0}\mathcal{L}_{v}(\vec{J}^{s}+\vec{J}^{p}), \label{eqDDA3}
\end{equation}
where $\eta_{0} = \sqrt{\frac{\mu_0}{\epsilon_0}}$ is the free-space impedance, $\epsilon_0$ and $\mu_0$ respectively denote the free-space electrical permittivity and magnetic permeability, 
$\vec{J}^{s}$ represents the independent current density, $\vec{J}^{p}$ is the induced current density, and the volumetric electrical field integral operator is \cite{YAGHJIAN1}


\begin{equation}
\mathcal{L}_v (\vec{J})= \left( -\jmath k_{0}\!\!\!\int_{V_{J}-V_{\delta}} \!\!\!\!\!\!\!\!\!\!\!\!{\bar{\bar{G}}}_{e}(\vec{r},\vec{r}') \cdot \vec{J}(r')dV'+\frac{\jmath}{k_0} {\bar{\bar{L}}} \cdot \vec{J}(\vec{r}) \right)
\label{eqDDA2}.
\end{equation}

In \autoref{eqDDA2}, $k \triangleq \omega \sqrt{\epsilon_0 \mu_0}$ is the free-space propagation constant and, $\vec{r}$ and $\vec{r}'$ are vectors indicating the source and observation 
points respectively. Moreover, $V_J$ denotes the volume in which the current is nonzero while $V_\delta$ is an infinitesimal volume enclosing
the observation point. ${\bar{\bar{G}}}_{e}(r,r')$ is the generalized dyadic \emph{Green}'s function. The tensor $\bar{\bar{L}}$ accounts for the effect of the current that resides at the observation point. As explained in 
\emph{Yanghjian}'s article \cite{YAGHJIAN1}, $\bar{\bar{L}}$ is obtained in the limit where volume approaches zero.
In all equations we use $\jmath$ for imaginary numbers. 
In the infinitesimal dipole model, polarization sources can be represented by \emph{Dirac} delta distributions. 
When observing the field at a point away from the source, the electrical field can be easily obtained using the electrical field representation formula for nonsingular
cases
\begin{align}
\mathcal{L}_{v}(\vec{J}) &=\jmath k_{0}\int_{\Omega}\vec{J}(\vec{r}')g(R)dv' + \jmath\frac{1}{k}\nabla\int_{\Omega}\nabla'.\vec{J}(\vec{r}')g(R)dv'\nonumber\\
& - \jmath\frac{1}{k_{0}}\nabla\int_{\partial\Omega}\vec{J}(\vec{r}')g(R).\hat{n'}ds',\: R\triangleq\left|\vec{r}-\vec{r}'\right|,
\label{eqDDA122}
\end{align}
where $k_0 = \omega \sqrt{ \epsilon_0 \mu_0}$ is the free-space wave number.
On the contrary, special care must be practiced when observing the field at points coinciding with the sources.
From \emph{Yanghjian}'s derivation \cite{YAGHJIAN1} the electric field at all points including those coinciding with the location of source currents
is obtained via the field operator $\mathcal{L}_v$ of \autoref{eqDDA2}. In \emph{Yaghjian}'s derivation \cite{YAGHJIAN1}, finite current densities are assumed. However, a critical difference here is that \emph{Dirac} delta distributions must be substituted into the derivation. In order to achieve this effect, one may assume $V_{\delta}$ with a uniform current density that is 
proportional to the inverse of the volume while the limit of the volume is approaching zero. In \autoref{eqDDA2}, the first term on right hand side is a nonsingular integral involving the conventional
dyadic $\bar{\bar{G}}_{e}$ outside the singularity region and the second term is the source dyadic $\bar{\bar{L}}$ which is determined solely from the geometry
of the principal volume chosen to exclude the singularity of $\bar{\bar{G}}_{e}$. In our case, the principal volume $V_\delta$ will be described  as a circular cylinder
in which both the radius and volume go to zero. Therefore, the first term in right hand side of \autoref{eqDDA2} is zero. On the other hand, 
it is assumed that $\int_{V_{\delta}}\!\!\vec{J}^{p}(\vec{r}')dV'$ remains constant while the dimensions of the principal volume $V_\delta$ approach zero. This
is equivalent to the assumption of a \emph{Dirac} delta distribution $\vec{J}^{p}\!\!=\!\!J^p\hat{s}\delta(\vec{r}')$. Under this assumption it is clear that
a finite source dyadic $\bar{\bar{L}}$ will not be obtained as in the case of \emph{Yanghjian}'s assumption \cite{YAGHJIAN1}.
Now, according to \cite{YAGHJIAN1}, the source dyadic of this principal volume is $\bar{\bar{L}}=\frac{1}{2}\bar{\bar{I}}_{t}$ and only has components orthogonal to the axis of the cylinder.
However, despite the infinite value of the field at the source, the orientation of the field will not be changed compared to that of \emph{Yanghjian}'s derivation.
Thus, if this field is tested by a vector quantity that is oriented along the axis of the cylindrical dipole, the
second term in \autoref{eqDDA2} will vanish. Hence, in the final discretized (matrix) equation which is due to the testing (collocation) of the field at the location of individual dipoles, 
the effect of the self term, i.e. the term due to testing of a dipole's field at the dipole itself, will not be present.
On the other hand, recall from \cite{JacksonBook} that for a dipole source $ \int \vec{J} dV = \jmath \omega \vec{p}$ which implies $\vec{J} = \jmath \omega \vec{p} \delta(r)$
for an infinitesimal dipole. Thus, using \autoref{eqDDA122}, the electric field due to a dipole located at points other than the dipole itself is
\begin{equation}
\vec{E}^{p}(\vec{r})\!\!=\!\!\frac{p}{4\pi\epsilon_{0}}\!\!\left(\frac{\left(1+\jmath k r\right)}{r^{2}}\left(3\hat{r}\left(\hat{r}.\hat{s}\right)\!-\!\hat{s}\right)-k^{2}\hat{r}\!\times\!\left(\hat{r}\!\!\times\!\!\hat{s}\right)\right)\!g(r)
\label{eqDDA7},
\end{equation}
where $p$ is the complex valued polarization intensity, $\hat{s}$ is a unit vector denoting the orientation of the dipole and $g(r)=\frac{e^{-\jmath kr}}{r}$ is the scalar free-space \emph{Green}'s function.

Now, consider a group of dipoles where the electric field of each dipole
can be expressed as \autoref{eqDDA7}. Considering a linear regime, the resulting polarization
$\vec{p}$ is equal to
\begin{equation}
\vec{p}(\omega)=\epsilon_{0} \bar{\bar{\chi}}(\omega)\vec{E}(\omega)\label{eqDDA8},
\end{equation}
where $\vec{E}$ is the (total) electric field on a dipole embedded in a medium of dielectric constant $\epsilon_{0}$ and $\vec{p}$ is the polarization of the dipole.
Thus, splitting the total electric field into internal (due to dipoles themselves) and incident field components, for the $m^{\textnormal{th}}$ dipole one writes
\begin{align}
\vec{p}_{m}&=\epsilon_{0} \bar{\bar{\chi}}_{m} \vec{E}^{inc}(\vec{r})|_{\vec{r}=\vec{r}_{m}} \nonumber \\
&\!+\!\epsilon_{0} \bar{\bar{\chi}}_{m} \!\!\sum_{n=1}^{N}\!\!\frac{p_{n}}{4\pi\epsilon_{0}}\left({\left(1\!+\!\jmath kr_{nm}\right)}\left(3\hat{r}_{nm}\left(\hat{r}_{nm}.\hat{s}_{n}\right)\!-\!\hat{s}_{n}\right)\!/\!{r_{nm}^{2}} \right. \nonumber \\
&\left.-k^{2}\hat{r}_{nm}\!\!\times\!\!\left(\hat{r}_{nm}\!\!\times\!\!\hat{s}_{n}\right)\right)g(r_{nm}), \label{eqDDA12}
\end{align}
where $\vec{r}_m$ is the location of the $m^{\textnormal{th}}$ dipole, 
$r_{mn}\triangleq|\vec{r}_n-\vec{r}_{m}|$ is the \emph{Euclidean} distance between the $m^{\textnormal{th}}$ and the $n^{\textnormal{th}}$ dipole,
$\hat{r}_{mn} \triangleq \frac{\vec{r}_n-\vec{r}_{m}}{|r_n-r_{m}|}$ and
$\hat{s}_{m}$ is the unit vector denoting the orientation of the $m^{\textnormal{th}}$ dipole. Moreover, one observes that the term involving $r_{nn}$ in \autoref{eqDDA12} will be infinite.
However as discussed earlier, by choosing a vector testing function that is oriented
along the $\hat{s}_{m}$, the effect of dipole electric field on itself will be dropped from the equation.
Here, a testing function of the form $\vec{E}_{t}=\delta\left(\vec{r}-\vec{r}_{m}\right)\hat{s}_{m}$ is used as an electric
field testing function where $\delta(\vec{r})$ is a $3\textnormal{-dimensional}$ \emph{Dirac} delta function satisfying 
\begin{equation}
\int_{\Omega} f(\vec{r}) \delta(\vec{r}-\vec{r}_0) dV = 
\left\{
\begin{array}{lr}
f(\vec{r}_0) & \vec{r}_0 \in \Omega\\
0 & \textnormal{otherwise}
\end{array}
\right..
\end{equation}
Hence, multiplying both sides of \autoref{eqDDA12} by $\vec{E}_{t}$
and integrating over the volume, one obtains the final discrete equation:

\begin{align}
p_{m}&=\epsilon_{0} \hat{s}_m \cdot \bar{\bar{\chi}}_{m} \vec{E}^{inc}(r)|_{r=r_{m}} +\nonumber \\
&\epsilon_{0} \hat{s}_m \!\!\cdot \bar{\bar{\chi}}_{m} \!\!\!\!\! \sum_{n=1, n \neq m}^{N}\!\!\!\!\frac{p_{n} }{4\pi\epsilon_{0}}\!\left(\!{\left(1\!\!+\!\!\jmath kr_{nm}\right)}\!\!\left(3\hat{r}_{nm}\!\!\left(\hat{r}_{nm}.\hat{s}_{n}\right)\!-\!\hat{s}_{n}\right)\!\!/\!{r_{nm}^{2}} \right. \nonumber \\
&\left.-k^{2}\hat{r}_{nm}\!\!\times\!\!\left(\hat{r}_{nm}\!\!\times\!\!\hat{s}_{n}\right)\right)g(r_{nm}). \label{eqDDA13}
\end{align}

\Autoref{eqDDA13} can be written in matrix form $\mathsf{Z}_{N\times N} \mathsf{P}_{N\times 1} = \mathsf{E}_{N\times 1}$ with the unknown vector $\mathsf{P} = [ p_{1},p_{2},\ldots,p_{N} ]^T$,
where $P_i$ are the complex values of the polarization of the individual dipoles located at $r_{1},r_{2},\ldots,r_{N}$.

\begin{align}
\mathsf{Z} = [\mathsf{Z}_{mn}], \mathsf{Z_{mn}} = 
\left\{
\begin{array}{lr}
1 & \textnormal{ if } m = n \\
\mathsf{Q}_{mn} & \textnormal{ otherwise }
\end{array}
\right.
\end{align}

\begin{align}
& \mathsf{Q}_{mn} \triangleq  -\epsilon_{0}  \frac{\hat{s}_m \!\cdot\! \bar{\bar{\chi}}_m }{4\pi\epsilon_{0}}  g(r_{nm}) \nonumber \\
&\!\left(\!\!\frac{\left(1\!\!+\!\!\jmath kr_{nm}\right)\!\! \left(3\hat{r}_{nm}\!\!\left(\hat{r}_{nm}.\hat{s}_{n}\right)\!-\!\hat{s}_{n}\right)}{r_{nm}^{2}} 
-k^{2}\hat{r}_{nm}\!\!\times\!\!\left(\hat{r}_{nm}\!\!\times\!\!\hat{s}_{n}\right)\!\!  \right)
\end{align}

The quantum mechanical properties of the monomers can be lumped into the tensorial complex response function $\bar{\bar{\chi}}(\omega)$.
In a general form, $\bar{\bar{\chi}}(\omega)$ is defined as 
\begin{equation}
\bar{\bar{\chi}}(\omega)\!=\!\frac{1}{\epsilon_{0}\hbar} \! \sum_{k}\!\!\frac{2\omega_{k} |{\rho_{0,k}|^2}}{\left(\omega\!+\!\omega_{k}\!+\!\jmath\frac{\gamma_{k}}{2}\right)\!\!\left(\omega\!-\!\omega_{k}\!+\!\jmath\frac{\gamma_{k}}{2}\right)}
\hat{u}_k\hat{u}_k\cdot
\label{eqQuantum1},
\end{equation}
where the sum over $k$ runs over all excited states of the monomer, $\omega_{k}$ is the transition frequency, $\gamma_{k}$ is the radiative decay rate, $\rho_{0,k}$ is the
transition dipole moment and $\hat{u}_k$ is a unit direction vector associated with the transition dipole moment between the ground state and that of the $k^\textnormal{th}$ state.
The $\omega_{k}$, $\gamma_{k}$ are considered to model the monomer as it is in its surrounding environment. This complex response function is nothing but the response of the monomer to a delta function electric field excitations in the time domain. In \autoref{eqQuantum1}, the $\omega_{k}$ in the denominator is associated with the monomers' transition energy
between the ground and $k^{\textnormal{th}}$ state while $\gamma_{k}$ is the so-called dephasing time. In \autoref{eqQuantum1},
as we will only consider the transition between the ground state and the first excited state of each monomer. The dipole moment unit vector $\hat{u}_1$ 
coincides with the physical orientation of the dipole $\hat{s}$ or $\hat{s}_m$ where $m$ labels the $m^{\textnormal{th}}$ monomer in the problem.

\section{Hierarchical Matrices}\label{introduce-HM}

\begin{figure}[TH]
\includegraphics[width=0.4\textwidth]{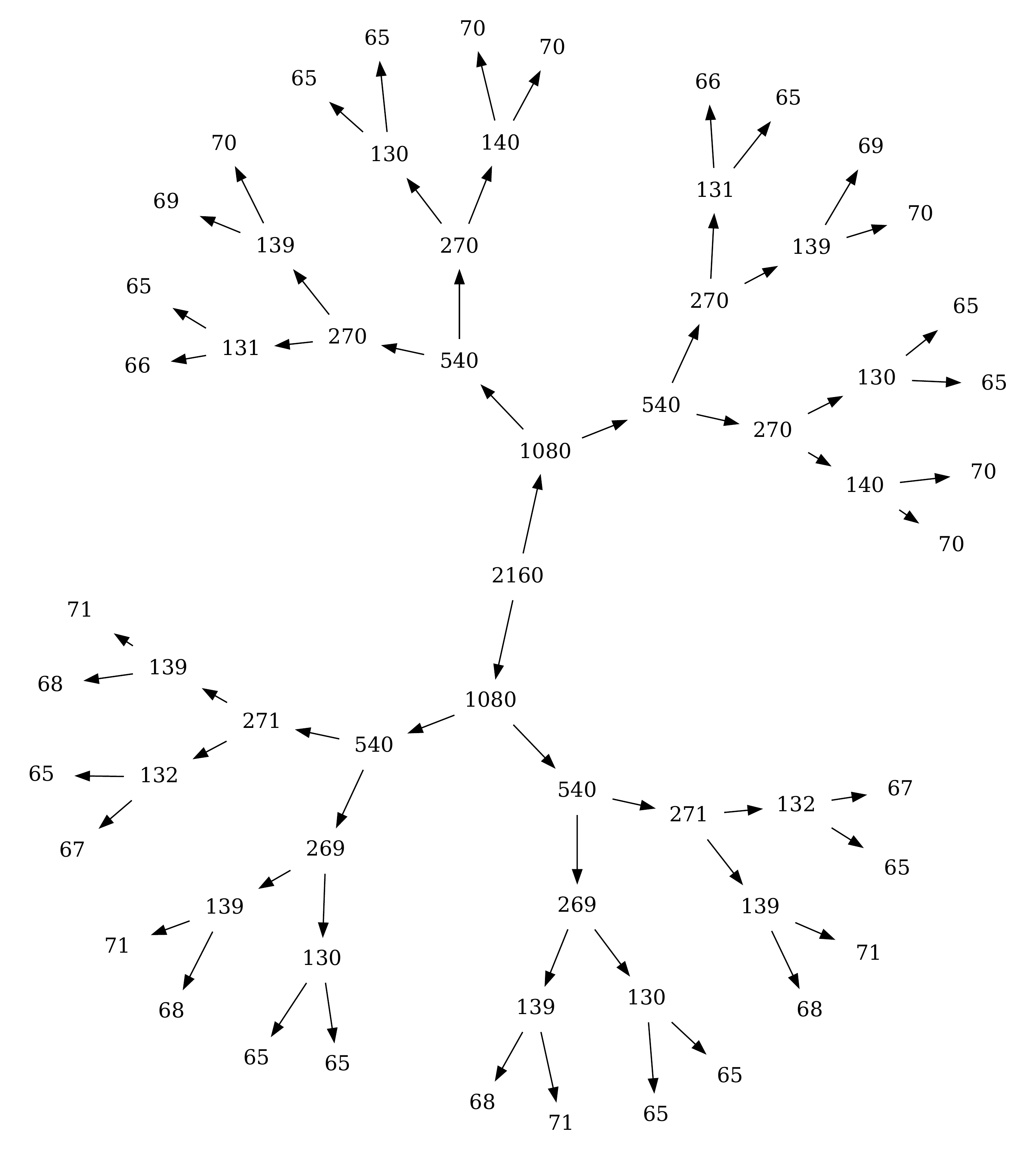}%
\caption{DoF cluster tree for a small hierarchical matrix corresponding to roll A consisting of 2160 monomers.
The numbers in the tree indicate the number of DoF associated with each node of the tree.
Each node in the tree represents a cluster of molecules (also DoF) in the problem structure.}
\label{clusterTreePlot}
\end{figure}

\begin{figure}[TH]
\includegraphics[width=0.4\textwidth]{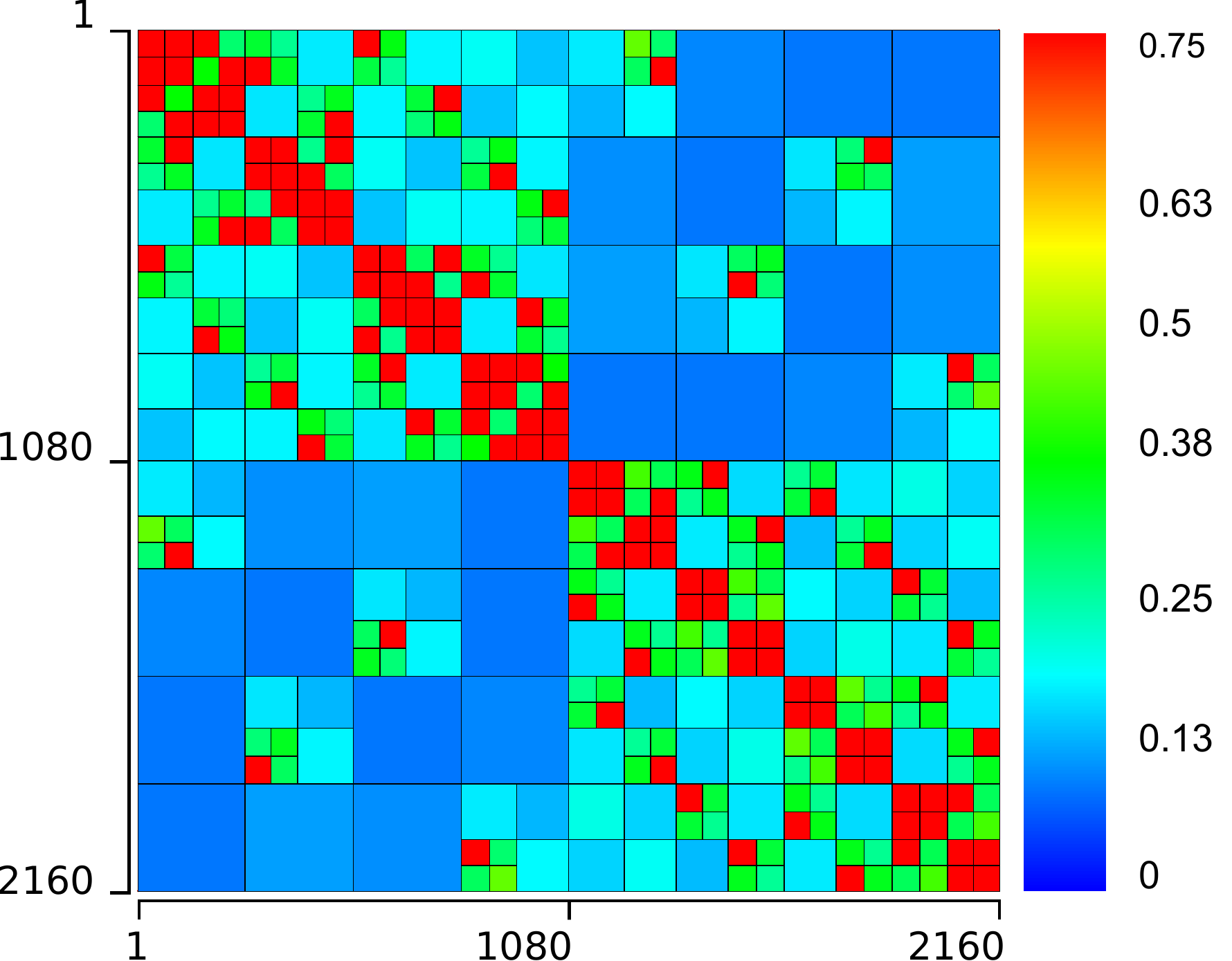}%
\caption{Compression ratio, defined as $k\frac{m+n}{m n}$ plotted for a relatively small hierarchical matrix corresponding to roll A consisting of 2160 monomers.}
\label{compressionPlot}
\end{figure}

Hierarchical matrices were originally applied to reduce the computational complexity associated with the discretization of elliptic differential equations \cite{HMatrix1, HMatrix2}.
It has been shown that for matrices resulting from the discretization of elliptic partial differential equations the \hm method leads to the almost 
linear complexity of $\mathcal{O}(N \log(N))$ \cite{HMatrix1, HMatrix2, HMatrix3} which is clearly advantageous over the $\mathcal{O}(N^2)$ complexity arising 
from the direct storage of the matrices. Although, the favorable properties of the \hm method will be lost if it is applied to non-elliptic operators, 
the method still provides very desirable performance in the so-called low frequency regime, i.e. when dimensions of the physical problem are smaller or comparable to the wavelength\cite{HMatrixHelmholtz1}.
Owing to their submicron dimensions, many molecular aggregates give rise to problems that exactly fit in to the low-frequency category.
The details of the \hm method can be found in \cite{HMatrix1,HMatrix2,HMatrix3,HMatrix4} and will not be discussed here.
Here it suffices to mention that in the \hm method, matrices are hierarchically reordered and segmented into blocks that are
classified into near-field and far-field interaction blocks.
The resulting near-field interaction blocks are stored as regular dense matrices while the far-field interaction blocks are stored via low rank
representation. In other words, a far-field interaction block $\mathsf{M}_{m\times n}$ is decomposed into a truncated SVD-like decomposition $\mathsf{M \approx U \Sigma V}^\dag$
in which singular values below a certain accuracy threshold are discarded. In practice, the low rank representation consists of a left factor
$\mathsf{A}_{m\times k}$ and a right factor $\mathsf{B}_{n\times k}$ satisfying $\mathsf{M \approx A B}^\dag$ where $k \ll \frac{m n}{m+n}$.
The fact that $k \ll \frac{m n}{m+n}$ guarantees that storage and other operations such as matrix-vector multiplication operations can be handled with a significantly reduced cost compared to that of regular dense matrix representation\cite{HMatrix1}. An example of a hierarchical subdivision of DoF and the corresponding
hierarchical matrix is depicted in \autoref{clusterTreePlot} and \autoref{compressionPlot} respectively. In \autoref{compressionPlot} both the structure and
the achieved compression ratio defined as $k \frac{m+n}{m n}$ can be observed. As can be seen, the near diagonal part of the matrix converts most of the self 
and near interactions and has the lowest levels of compression.

The time domain solution of the problem can be obtained by applying FFT to a large collection of frequency dependent solutions.
Considering that a small number of \emph{Krylov} iterations are needed for the problem solution at each frequency, most of the computational cost
will be associated with the construction of the \hmp. Hence, one may naturally think of interpolating
the majority of these frequency dependent \hmp from a limited set generated at a  selected frequencies.
Note that here the term `interpolation' is used in its loose sense and shall imply the meaning of any curve fitting or parameterization technique.
In the following section, an efficient parameterization method for \hmp is proposed.

\section{Parametric \hm Representation}\label{OptimalRep}

Let us consider a block of the system matrix ${\mathsf{Z}}_{(m \times n)}$.
We intend to build a frequency dependent characterization of the complete system matrix.
The characteristic to our problem of interest, $\mathsf{Z}(f)$, is a smooth and infinitely differentiable function, except near the frequency points in the vicinity of
the quantum mechanical resonances of individual monomers as discussed in \autoref{DDAFrom}. Due, to the presence of such singularities (poles) 
in the frequency representation of individual matrix entries, polynomial interpolation or polynomial least squares curve fitting methods are not applicable. 
Nonetheless, a rational function representation can effectively represent the individual matrix entries. Such a rational function representation can be obtained using the \emph{vector fitting} 
method \cite{VF1,VF2}. Thus, if the system matrix is represented in a dense form, the entries can be fitted to rational functions by means of the abovementioned method.
Nevertheless, for large systems the computational complexity of the dense matrix representation is prohibitive.

Alternatively, in the \hm representation, two types of blocks are present: (1) dense blocks in the form of $\mathsf{M}$ used for near interactions 
(2) low rank (compressed) blocks $\mathsf{M} = \mathsf{A} \mathsf{B}^\dag$. The dense blocks can be fitted directly into a rational function of frequency using the
\emph{vector fitting} method. On the other hand, for the low rank blocks, the frequency parameterization cannot be directly applied to the $\mathsf{A}$ and $\mathsf{B}^\dag$ factors.
In other words, direct interpolation/parameterization of $\mathsf{A}(f)$ and $\mathsf{B}(f)$ for the low rank blocks would result in undesirable results since the smoothness 
of $\mathsf{A}(f)$ and $\mathsf{B}(f)$ is not guaranteed, although the product $\mathsf{M} = \mathsf{A} \mathsf{B}^\dag$ is known to be sufficiently smooth. 
This is because the matrices $\mathsf{A}(f)$ and $\mathsf{B}(f)$ are not unique, i.e.  if $\mathsf{M} = \mathsf{A} \mathsf{B}^\dag$,
so will $\mathsf{M} = (\mathsf{A} \mathsf{C})(\mathsf{C}^{-1} \mathsf{B}^\dag)$ for any $\mathsf{C} \in \mathbb{C}^{k \times k}, \det\mathsf{C} \neq 0$, $\mathsf{C}^{\dagger} = \mathsf{C}^{-1}$.
Furthermore, redundancies may exist between the members of the sets $\{ \mathsf{A}_1, \mathsf{A}_2, \cdots
, \mathsf{A}_{n_f} \}$ and $\{ \mathsf{B}_1, \mathsf{B}_2, \cdots , \mathsf{B}_{n_f} \}$ that respectively represent the range and the domain spaces of the frequency dependent matrix ${\mathsf{M}}$
at a selection of frequencies $\{f_1,f_2,\ldots, f_{n_f}\}$.

Suppose $n_f$ sampling frequencies are given as the key data points for the intended parameterization and thus the low rank
representation of $\mathsf{M}$ using $\{\mathsf{A}_1, \ldots, \mathsf{A}_{n_f} \}$ and $\{\mathsf{B}_1, \ldots, \mathsf{B}_{n_f} \}$ is given. As one moves from $f_p$ to $f_q$, the range space of the operator
$\mathsf{M}$ changes from $\mathsf{A}_p$ to $\mathsf{A}_q$. However, due to the finite dimension and 
the smooth frequency dependence of operator $\mathsf{M}$, it is expected that the range spaces of $\{\mathsf{A}_1, \ldots, \mathsf{A}_{n_f} \}$ share common information. 
In order to extract the potentially existing redundancies in the set $\{\mathsf{A}_1, \ldots, \mathsf{A}_{n_f} \}$, the SVD can be applied as
\begin{align}
&\{ \mathsf{A}_{i,(m \times k_i)} \} \rightarrow  \nonumber \\
&[\mathsf{A}_1, \ldots, \mathsf{A}_n ]_{(m \times \sum k_i)} = \mathsf{U}_{L,(m \times m)} \mathsf{\Sigma}_L \mathsf{V}^\dag_{L,(\sum k_i
\times \sum k_i)},
\end{align}
where $\{ \mathsf{A}_{i,(m \times k_i)} \}$ is a set of range-space matrices and the subscript \textit{L} denotes its association with left factor $\mathsf{A}$ in the original decomposition
of $\mathsf{M}$. With a desired level of accuracy, the above SVD can be truncated and written as
\begin{equation}
[\mathsf{A}_1, \ldots, \mathsf{A}_n ] \approx \mathsf{\bar{U}}_{L,(m \times k_L)} \mathsf{\bar{V}}_{L, (k_L \times \sum k_i)}^\dag.
\end{equation}

Under the same truncation tolerance, each of the $\mathsf{A}_i$ matrices can be decomposed as $\mathsf{A}_{i,(m \times k_i)} = \mathsf{\bar{U}}_{L, m
\times k_L } \mathsf{V}_{L,i, (k_L \times k_i)}^\dag$, where $\mathsf{V}_{L,i}$ represents the portion of row vectors
in $\mathsf{\bar{V}}_L$ that corresponds to the construction of $\mathsf{A}_i$.
Applying a similar procedure to the right side factor, $\{\mathsf{B}\}_i$, we get
\begin{equation}
\left[
\begin{array}{c}
 \mathsf{B}^\dag_1 \\
 \vdots \\
 \mathsf{B}^\dag_n \\
\end{array}
\right]_{(\sum k_i \times n)} \approx \mathsf{\bar{U}}_{R, (\sum k_i \times k_R) } \mathsf{\bar{V}}_{R, (k_R \times n)}^\dag.
\end{equation}
Again, each $\mathsf{B}_{i}$ matrix is written as $\mathsf{B}_{i, (k_i \times n)}^\dag \approx \mathsf{\bar{U}}_{R,i,(k_i \times k_R)} \mathsf{\bar{V}}_{R, (k_R \times n)}^\dag
$,
where $\mathsf{\bar{U}}_{R,i}$ represents the collection of row vectors
in $\mathsf{\bar{U}}_R$ corresponding to the construction of $\mathsf{B}_i$.
Now, one can state that 
\begin{equation}
\mathsf{M}_i \approx \mathsf{\bar{U}}_L \mathsf{T}_i \mathsf{\bar{V}}_R^\dag,
\end{equation}
where
\begin{equation}
\mathsf{T}_{i, (k_L \times k_R)} \triangleq \mathsf{\bar{V}}_{L,i,(k_L \times k_i )}^\dag \mathsf{\bar{U}}_{R,i, (k_i \times k_R)}
\end{equation}
is only a frequency dependent part in the low rank representation of the block $\mathsf{M}_i$. The $\mathsf{T}_{i}$ matrix: (1) is unique for each frequency, and (2) it has
smaller dimensions compared to the matrices $\mathsf{M}_i$, $\mathsf{A}_i$ and $\mathsf{B}_i$. Thus, lending itself to more efficient computational operations.

\section{Error Control}\label{ErrCon}

Errors in the parametric hierarchical matrix representation can be divided into two main categories: (a) 
truncation errors and (b) parameterization errors, where the former are associated with the low rank representations used in the \hm 
and the latter correspond to the parameterization (interpolation or curve fitting) procedure. There are no truncation errors for dense blocks. 
Thus, only parameterization errors should be minimized.
For low rank blocks, however, some care must be taken to properly control both parameterization and truncation error while imposing minimal computational costs. 
For this purpose, let us consider the following parametric representation of a low rank block $\mathsf{M}(f)$
\begin{align}
\mathsf{M}(f) = \mathsf{U}_L \mathsf{\Sigma}_L \mathsf{T}(f) \mathsf{\Sigma}_R \mathsf{V}^\dag_R,
\end{align}
where we assume that no SVD truncation has been applied to the left and right factors $\mathsf{U}_L \mathsf{\Sigma}_L$ and
$\mathsf{\Sigma}_R \mathsf{V}^\dag_R$.

In order to model the truncation error, assume that $\mathsf{\Delta\Sigma}_L$ and $\mathsf{\Delta\Sigma}_R$ represent the part of the
singular value spectrum that is eventually removed due to the low rank representation.
Also, let's assume that a $\mathsf{\Delta T}$ error is introduced to the matrix $\mathsf{T}$ due to the parameterization. 
Then, the parameterized low rank representation is
\begin{align}
\bar{\mathsf{M}} = \mathsf{U}_L (\mathsf{\Sigma}_L-\mathsf{\Delta\Sigma}_L) (\mathsf{T}+\mathsf{\Delta T})
(\mathsf{\Sigma}_R-\mathsf{\Delta\Sigma}_R) \mathsf{V}^\dag_R.
\end{align}
Assuming that all three sources of error, i.e. $\mathsf{\Delta \Sigma}_L$, $\mathsf{\Delta \Sigma}_R$ and $\mathsf{\Delta
T}$ are small relative to their central values, and applying the \emph{Frobenius} norm as an error measure we can write
\begin{align}
\norm{e}_F \approx
\norm{
\-\mathsf{\Delta\Sigma}_L \mathsf{T} \mathsf{\Sigma}_R
-\mathsf{\Sigma}_L \mathsf{T} \mathsf{\Delta\Sigma}_R
+\mathsf{\Sigma}_L \mathsf{\Delta T} \mathsf{\Sigma}_R
}_F. \label{eqNorm}
\end{align}
In \autoref{eqNorm} only terms linear in the error, i.e. lowest order perturbations, are included and the unitary matrices, $\mathsf{U}$ and $\mathsf{V}^\dag$ are 
discarded due to the invariance of the \emph{Frobenius} norm upon a unitary transformation.  
The upper bound for the error, \autoref{eqNorm}, can be derived using the triangle inequality
\begin{align}
\norm{
-\mathsf{\Delta\Sigma}_L \mathsf{T} \mathsf{\Sigma}_R
-\mathsf{\Sigma}_L \mathsf{T} \mathsf{\Delta\Sigma}_R
+\mathsf{\Sigma}_L \mathsf{\Delta T} \mathsf{\Sigma}_R
}_F
\leq \nonumber \\
\norm{
\mathsf{\Delta\Sigma}_L \mathsf{T} \mathsf{\Sigma}_R
}_F
+
\norm{
\mathsf{\Sigma}_L \mathsf{T} \mathsf{\Delta \Sigma}_R
}_F
+
\norm{
\mathsf{\Sigma}_L \mathsf{\Delta T} \mathsf{\Sigma}_R
}_F.
\end{align}

The three terms in the above expression indicate that there are three sources of error, two related to the truncation of the left and the
right factors and one coming from the error caused by parametrization of matrix entries. The truncation error contributions can be controlled through proper
truncation of singular values in the left and right factors, pretty much as it is done for the low rank blocks in non-parametric hierarchical matrices. This 
bound serves as a convenient measure of the accuracy of the parametric $\mathcal{H} \textnormal{-matrix}$ blocks and can be used for assessment of the
success of the parametrization.

Moreover, the $\mathsf{\Sigma}_L \mathsf{\Delta T} \mathsf{\Sigma}_R$ directly reflects how the error due to the parameterization procedure is manipulated by 
the left and right factors, i.e. $\mathsf{\Sigma}_L$ and $\mathsf{\Sigma}_R$. The immediate consequence, however, is that not all entries in $\mathsf{T}$ need 
to have the same level parameterization error as these entries are scaled by the singular values in $\mathsf{\Sigma}_L$ and $\mathsf{\Sigma}_R$. 
Therefore, in order to control the parameterization error in $\mathsf{T}$,
one needs not to directly control the error in the individual entries $\mathsf{T}_{ij}$, but rather that of $\mathsf{\Sigma}_{L,ii}
\mathsf{T}_{ij} \mathsf{\Sigma}_{R,jj}$. In other words, the error introduced due to the parameterization of the $\mathsf{T}_{ij}$ entries with higher values of $i$ and $j$ is
less important as it will be multiplied by smaller singular values. In practice, this balances out with the more significant error level that is observed
in the parameterization error induced $\mathsf{T}_{ij}$ entries with higher values of $i$ and $j$. In this light, the individual entries of the $\mathsf{T}$
matrix can be observed as the modal functions from which the matrix block $\mathsf{M}$ is constructed as a function of frequency. Intuitively, one observes
that the modal functions with higher indices have more complicated behavior and thus are more difficult to parameterize.

\section{Chlorosome Roll Model}\label{Model}

\begin{figure}[TH]
  \subfigure[]
  {
    \includegraphics[width=0.15\textwidth]{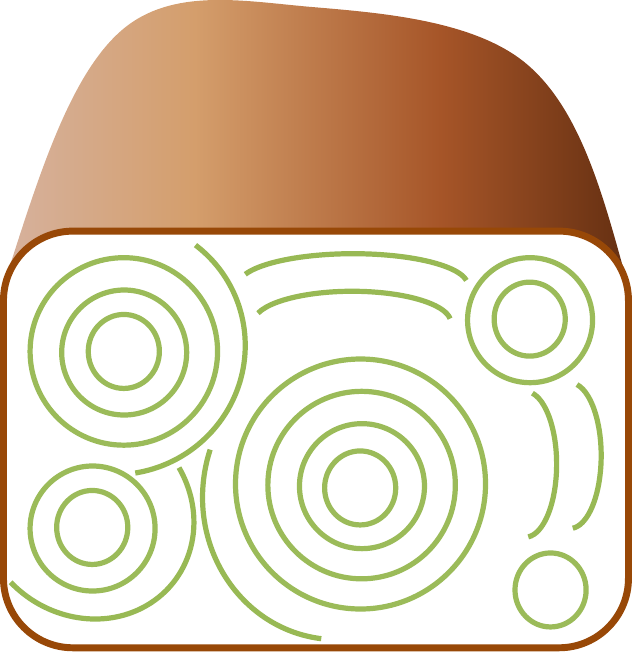}
  } 
  \subfigure[]
  {
    \includegraphics[width=0.2\textwidth]{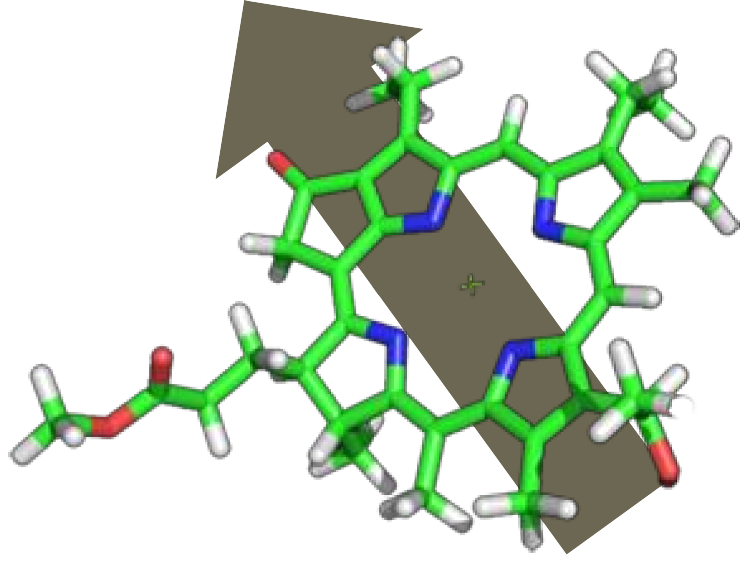}
  } \\
  \subfigure[]
  {
    \includegraphics[width=0.4\textwidth]{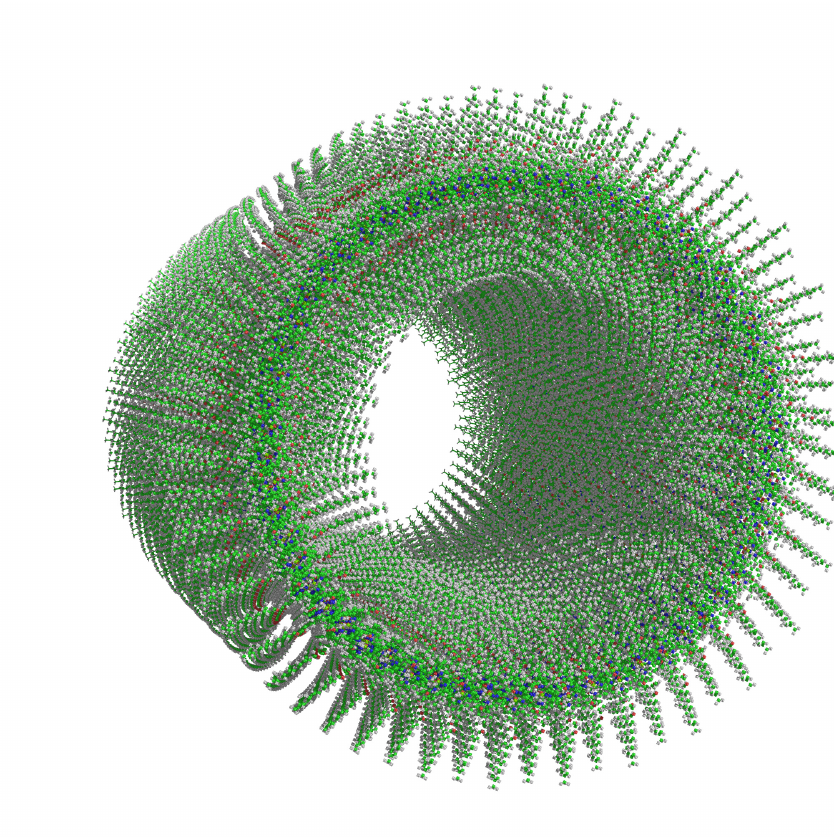}
  }   
  \caption{
           (a) A schematic structure of the chlorosome antenna in green sulfur bacteria. Tubular and curved lamella aggregates of pigment bacteriochlorophyll molecules form an 
           ovoid-shaped body with characteristic size of tens to hundreds of nanometers. (b) A bacteriochlorophyll molecule with the orientation of the lowest transition dipole 
           shown. (c) Packing of BChl transition dipoles in a roll aggregate forms a collection of infinitesimally small electrical dipoles.
          }
  \label{fig:chlorosome}
\end{figure}

It is understood that the chlorosome is composed of multiple rolls and curved lamella structures as schematically illustrated in \autoref{fig:chlorosome}(a). While there are 
several models for BChl packing in the aggregates~\cite{Holzwarth94,Pcencik2004,Ganapathy2009}, we use the one suggested recently by Ganapathy et. al. \cite{Ganapathy2009}. The lowest 
electronic excitation in single BChls, $Q_y$ band, is about 1.8~eV or 435~THz with the orientation of the transition dipole shown in \autoref{fig:chlorosome}(b). The $Q_y$ band is 
separated sufficiently from the next transition, $Q_x$ band. Thus, in our modeling only the lowest electronic excitation is considered. According to the model~\cite{Ganapathy2009} BChl pigment 
molecules are arranged in concentric rings and then stacks of these rings form a roll. The molecular transition dipoles are almost orthogonal to the radius and form 35 degrees angle with the 
plane of the ring. 
\Autoref{fig:chlorosome} (c) schematically depicts the molecular structure of one of the rolls consisting of the bacteriochlorophyll molecules depicted in \autoref{fig:chlorosome}(b).

To model the chlorosome response four concentric rolls A, B, C and D with $N_{A-D} = 60,80,100,120$ number of molecules per ring (see \autoref{probes} for details), 
and $n = 36$ rings were constructed and their structure was compared to the structures provided to us by the authors of reference \cite{Ganapathy2009}. 
The molecular structure depicted in \autoref{fig:chlorosome} (c) corresponds to the innermost roll A.
For transient response calculations, we selected six points (and monomers) $a-f$ coinciding some of the monomers in the four-roll structure. Points $a$,$c$ ($d$,$f$) are located 
on the edges of the inner (outer) rolls, and points $b$ and $e$ belong to the central rings of the same rolls, as shown in \autoref{probes}. Clearly, the line shifts associated with the retardation cannot be observed at ambient conditions, where the resonance lines are broadened by about 50 meV due to the structural disorder and thermal effects. However, the role of disorder in the intensity redistribution cannot be easily analysed and should be studied in more detail.

We illustrate the developed computational method by modeling the optical responses of BChl roll aggregates, contained in the chlorosome antenna complex of green photosynthetic bacteria 
\cite{Oostergetel}. All results presented in this article were obtained using a C++ code compiled with the GNU C++ version $4.6.3$ compiler on a Linux based dual 
6-core \emph{Intel Xeon 5649} workstation although the muti-core features of the machine were not used in our current implementation. 
All reported timings are based on single-thread runs without parallelization.

\subsection{Spectral analysis}

\begin{figure}[TH]
  \subfigure[Semiclassical model]
  {
    \label{Semiclassical}
    \includegraphics[width=0.475\textwidth]{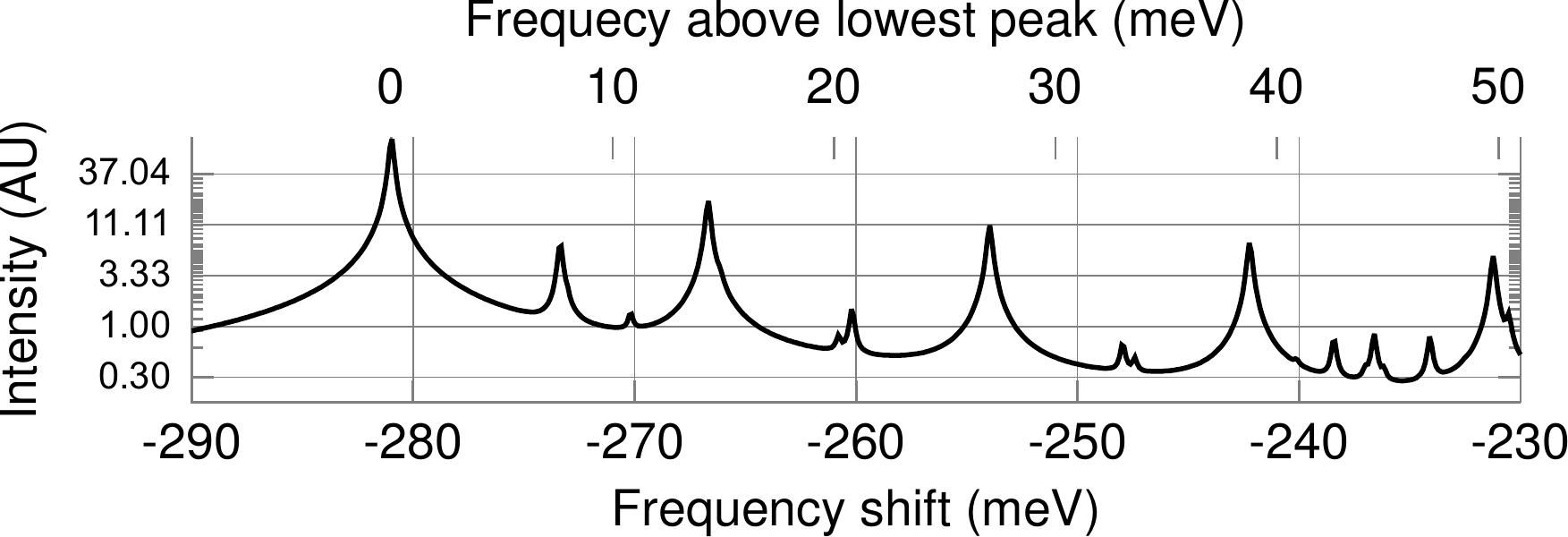}
  } \\
  \subfigure[Quantum model]
  {
    \label{Quantum}
    \includegraphics[width=0.45\textwidth]{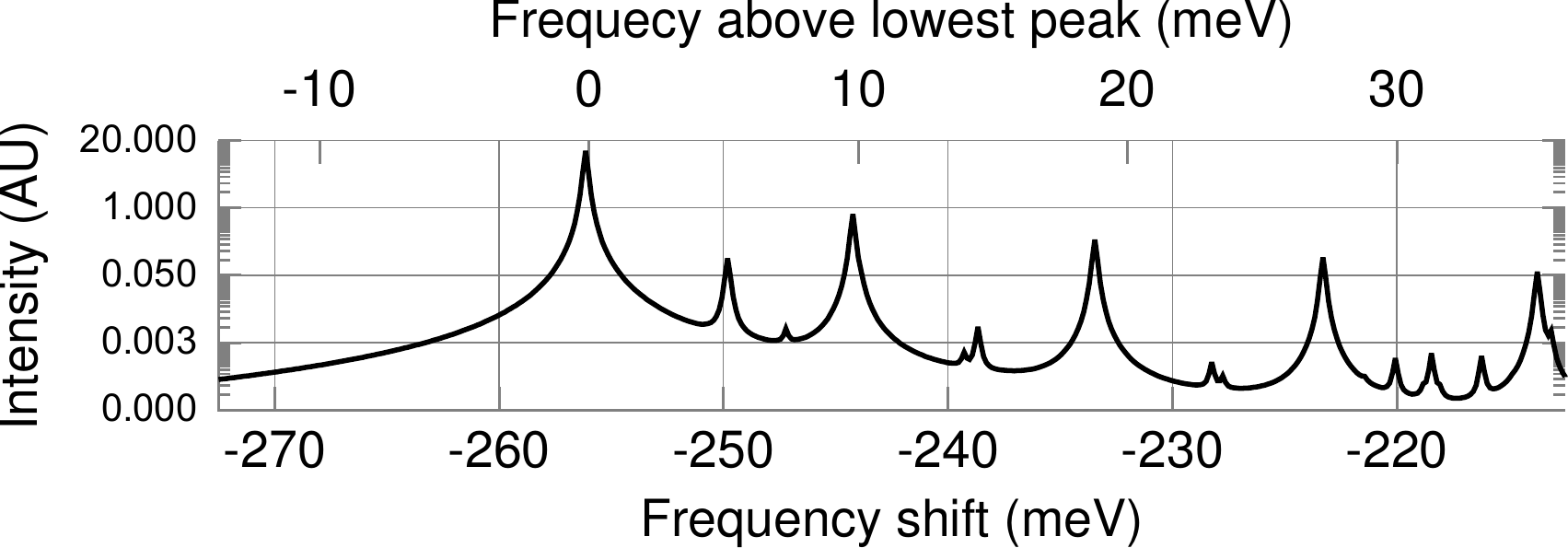}
  }
  \caption{Frequency spectra of roll D in response to $\hat{z}$-polarized optical field computed using a quantum mechanical and a hybrid electrodynamic method.
The numbers on the bottom horizontal axis indicate the relative red shift in $\textnormal{meV}$ with respect to the resonant frequency of the individual
molecules located at $1.8\textnormal{eV}$ which is equivalent to $435\textnormal{THz}$. 
In the semicalssical results, the vertical axis shows the sum of the squared polarization amplitudes of individual molecules. To enhance the details the vertical axis is in log scale.}
  \label{spectrumCompareZ}
\end{figure}

As a first step we compared the resonance spectra of roll D in Fig. 9 simulated using the hybrid quantum-classical formulation with that of the quantum Hamiltonian model \cite{Taka2012} calculated as described in reference\cite{Somsen1996}.
The hybrid model used in this article, takes the quantum mechanical effects into account via the polarizability factor of individual monomers, while the interaction 
between molecules is considered within a self-consistent linear response theory. \Autoref{spectrumCompareZ} shows the spectra computed with both models, where the resonance frequency of isolated 
monomers $1.8$~eV was used as a baseline and the linewidth was assumed to be 0.1~meV in order to resolve different resonances. The resonances of the chlorosome aggregates are red-shifted from 
the monomer frequency in both quantum and hybrid model in agreement with the experimental data~\cite{Oostergetel}. Apart from a frequency shift attributed to the difference between the quantum and the classical model 
(and also due to the inclusion of retardation effects since the quantum model only accounts for the electrostatic interactions) the structures of the computed spectra are very similar. Please note that the difference in the spectra obtained by means of these models cannot be seen as a perfect red shift since larger shifts are observed at lower frequencies.

\begin{figure}[TH]
\includegraphics[width=0.475\textwidth]{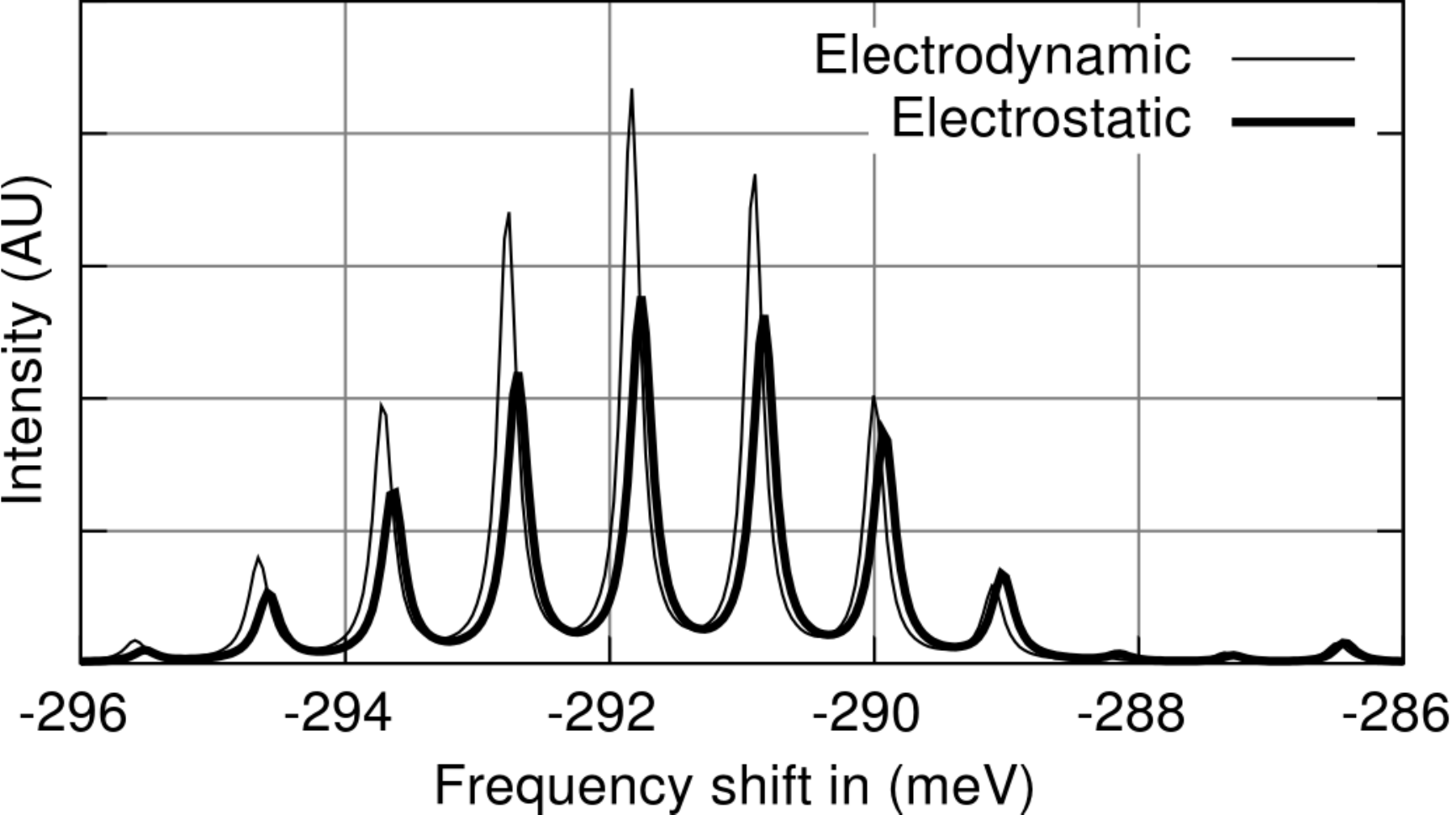}%
\caption{
Frequency spectrum of the extended roll D in response to $\hat{x}$-polarized initial electric field using both electrostatic and electrodynamic formulation.
The numbers on the bottom horizontal axis indicate the relative red shift in $\textnormal{meV}$ with respect to the resonant frequency of the individual
molecules located at $1.8\textnormal{eV}$.
The vertical axis shows the sum of the squared polarization amplitudes of individual molecules. To enhance the details the vertical axis is in log scale.
}%
\label{dynamic-static-Xpol}
\end{figure}

\begin{figure}[TH]
\includegraphics[width=0.46\textwidth]{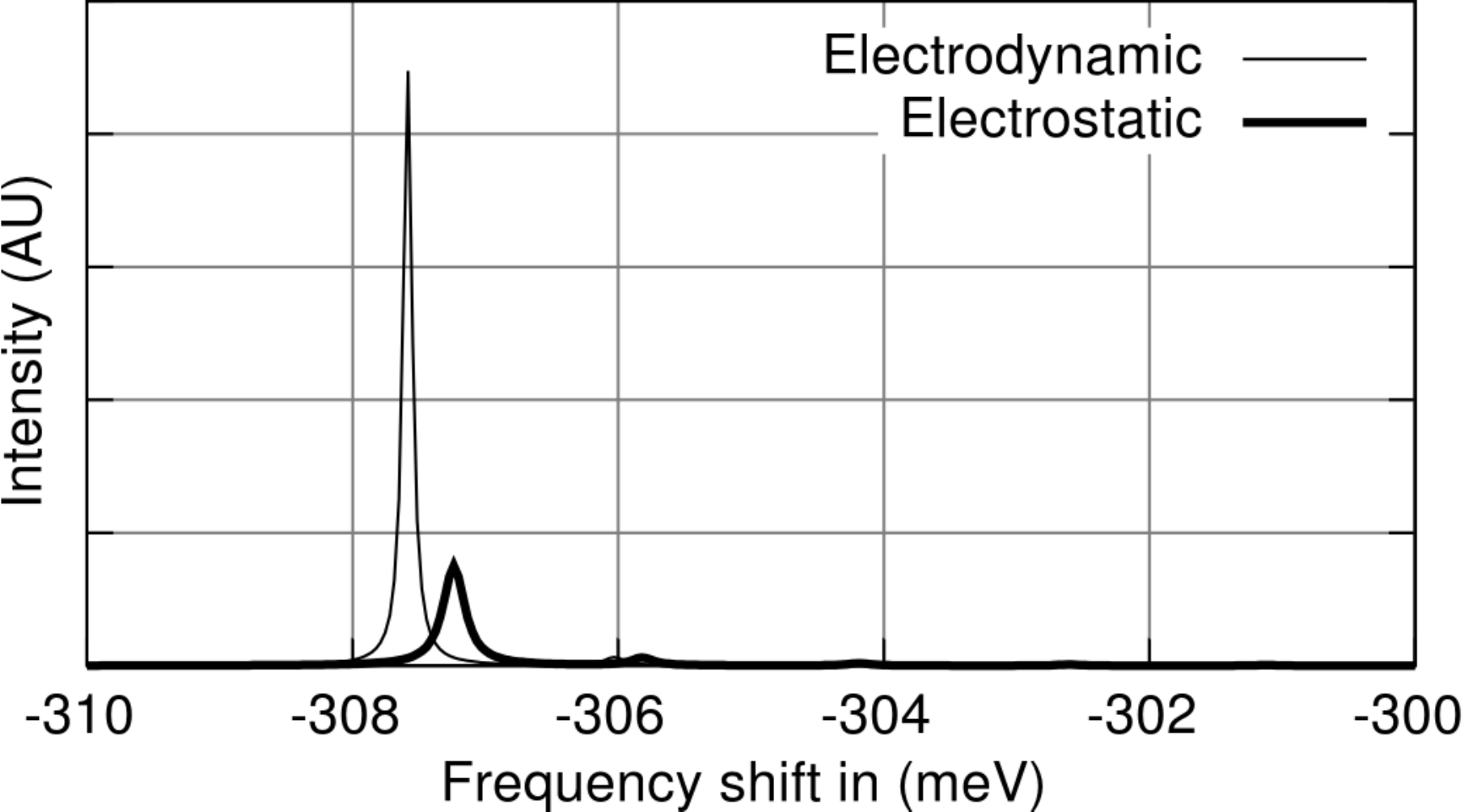}%
\caption{
Frequency spectrum of the extended roll D in response to $\hat{z}$-polarized initial electric field using both electrostatic  and electrodynamic formulation.
The numbers on the bottom horizontal axis indicate the relative red shift in $\textnormal{meV}$ with respect to the resonant frequency of the individual
molecules located at $1.8\textnormal{eV}$. 
The vertical axis shows the sum of the squared polarization amplitudes of individual molecules.
}%
\label{dynamic-static-Zpol}
\end{figure}

Both electrodynamic and electrostatic formulations have been proposed previously to study the interaction of light with molecular aggregates 
\cite{DeVoe_JCP1965,Keller_JCP1986, DDA1}. Intuitively, when the physical dimension of the structure is much smaller than the wavelength of light, 
electrodynamic retardation effects are expected to be insignificant. However, as the dimensions of the structure increase and become comparable with the wavelength,
those effects can become more pronounced. For this purpose, we constructed an long roll that has the same radius as roll D but length $290\textnormal{ nm}$, which is comparable to the physical length of a chlorosome~\cite{Oostergetel}. The resulting 
structure for this `extended roll D'  consists of $N = 43200$ molecules. The obtained spectra, \autoref{dynamic-static-Xpol} and \autoref{dynamic-static-Zpol}, show that the retardation effects 
can result in a sufficient redistribution of peak intensities within the aggregate especially for the fields polarized along the roll's symmetry axis.

\begin{figure}[TH]
\includegraphics[width=0.46\textwidth]{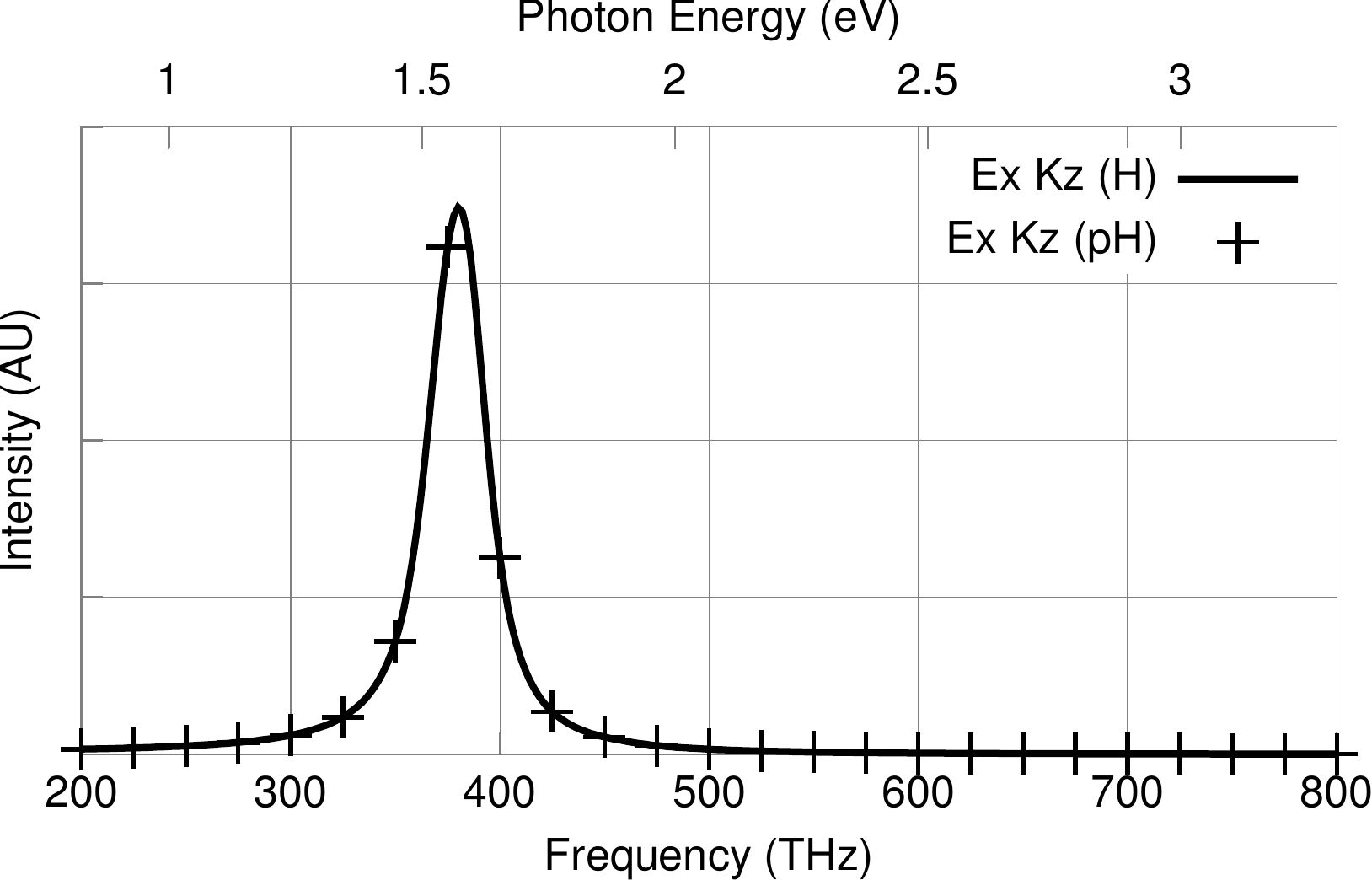}%
\caption{
Response to plane wave excitation traveling along $\hat{z}$ and
the electrical field polarized along $\hat{x}$.
In the legend, `H' and `pH' indicate \hm and  parametric \hm method respectively.
The vertical axis shows the sum of the squared polarization amplitudes of individual molecules.
\label{fig1}
}%
\end{figure}

\begin{figure}[TH]
\includegraphics[width=0.46\textwidth]{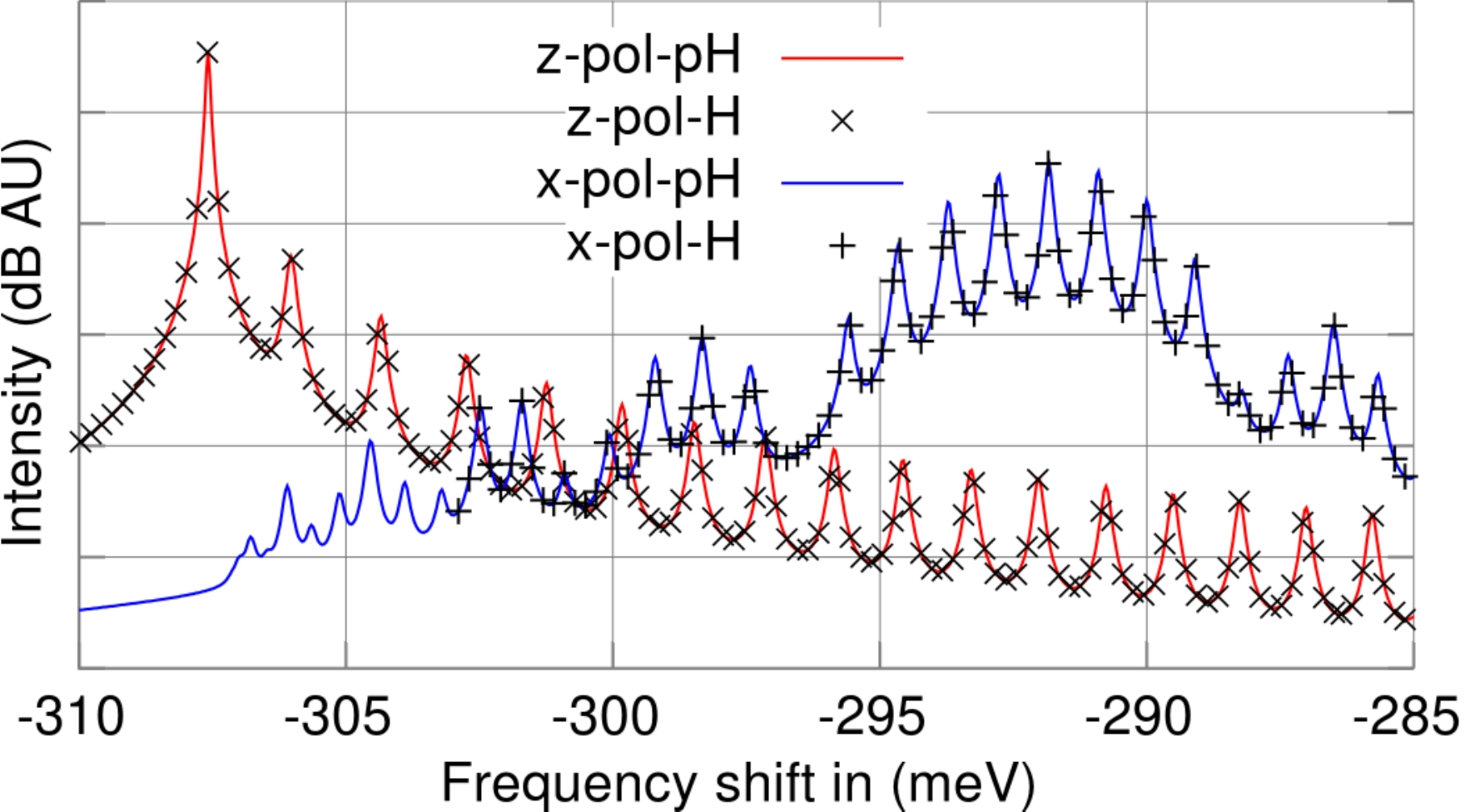}%
\caption{\label{slow-fast} Comparison of the spectral response of extended roll D structure using both
\hm and parametric \hm method. The response is due to initial field excitation along $\hat{z}$ direction.
The vertical axis shows the sum of the squared polarization amplitudes of individual molecules.
To enhance the details the vertical axis is in log scale.
}
\end{figure}

To examine the reliability of the parametric \hm method two example problems were solved using both the parametric \hm and the nonparametric \hm method. 
As a first example the bacteriochlorophyll roll of \autoref{probes} was excited with a steady state plane wave and polarization intensities were calculated as a function of the 
excitation frequency. The excitation decay rate was assumed to be the same for all monomer with equal $\gamma = 50$~meV. \Autoref{fig1} shows a quantitative agreement between 
the response obtained by means of the parametric \hm method at 300 frequencies (in the 200 THz to 800 THz band) and twenty five sample 
points obtained using the \hm method. As a second example a similar comparison was done for the $290$~nm long roll, see \autoref{slow-fast}, 
which again showed a perfect agreement between the results.

\subsection{Transient Response of Roll Aggregates}\label{TSingle}

\begin{figure}[TH]
\includegraphics[width=0.3\textwidth]{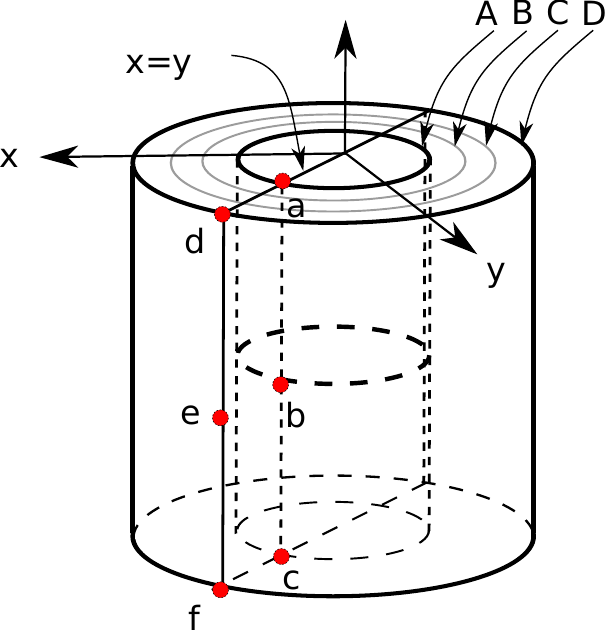}
\caption
{
Location of observation monomers in the four-layer chlorosome roll. The letters a, b, c, d, e and f indicate the six 
monomers used for the observation of the time evolution of their polarization amplitude as depicted in \autoref{singleTrans}.
The capital letters A, B, C and D correspond to the four layers discussed in \autoref{Model}.
}\label{probes}
\end{figure}

Here, the chlorosome roll of \autoref{probes} is excited with initial $\hat{z}$ polarized electric field incident on monomers 
located at the central ring of roll A, i.e. the innermost roll as illustrated in \autoref{probes}, and the time evolution of the polarization induced in various monomers at the 
points $a-f$ in the four-layer roll, see \autoref{probes}, is observed. The time domain is constructed from a large number of frequency domain solution via FFT. Thus it is natural 
to use the parametric \hm method to rapidly produce the required frequency response. The results presented in this section were obtained via $350$~frequency 
domain solutions covering the frequency band from 0 THz to 800 THz. The parametric \hm is produced from $13$ \hmp evenly covering the 
100 THz to 700~THz band. Generation of each \hm takes about $172$~seconds ($2236$~seconds total) and the construction of the parametric \hm 
takes another $2000$~seconds. However, when the parametric \hm is ready, the construction of each \hm at
a given frequency only requires $7$~seconds. At most frequencies, very few ($4$ to $50$) \emph{Krylov} iterations are needed for the matrix solution process and thus the computational time is mainly determined by the 
time spent on the construction of the system \hm. Under this configuration, the parametric \hm method leads to an overall speedup factor of $8$ compared to the direct use of the \hm method. 
It is worth mentioning that higher speedup factors will be achieved when a larger number of frequency domain solutions are required or when larger structures need to be solved.

\begin{figure}[TH]
     \begin{center}
        \subfigure[@ a]
        {
            \includegraphics[width=0.22\textwidth]{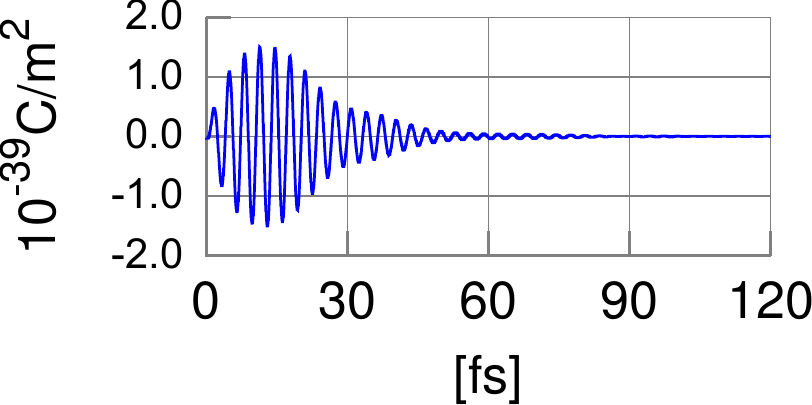}
        } \hfill
        \subfigure[@ d]
        {
            \includegraphics[width=0.22\textwidth]{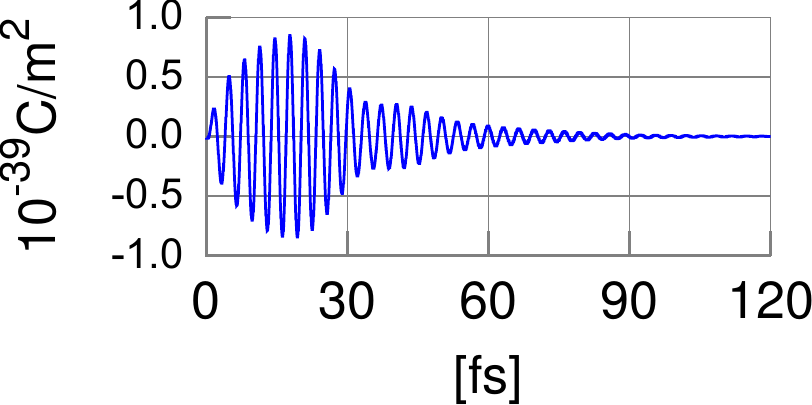}
        } \\
        \subfigure[@ b]
        {
            \includegraphics[width=0.22\textwidth]{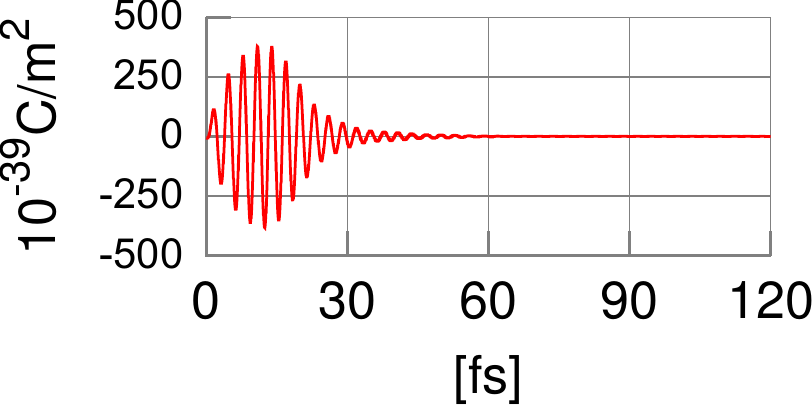}
        } \hfill
        \subfigure[@ e]
        {
           \includegraphics[width=0.22\textwidth]{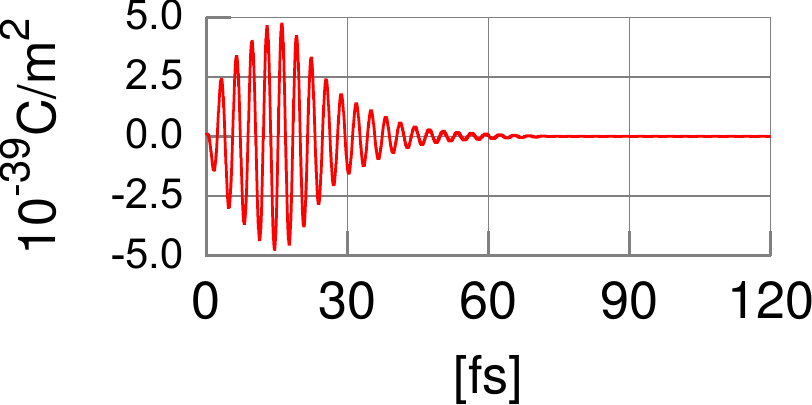}
        } \\
        \subfigure[@ c]
        {
           \includegraphics[width=0.22\textwidth]{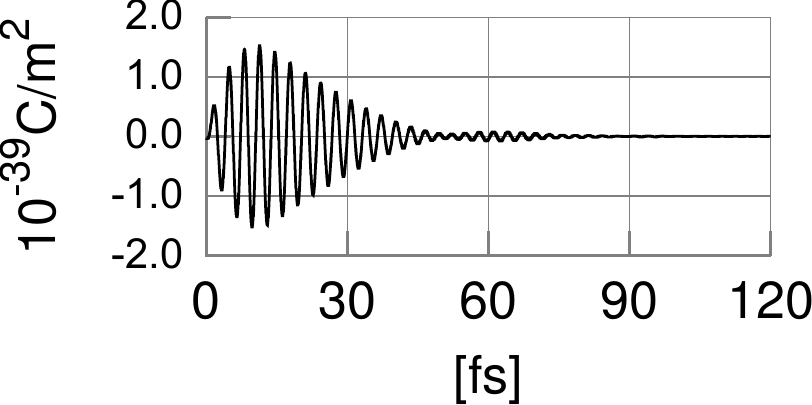}
        } \hfill
        \subfigure[@ f]
        {
           \includegraphics[width=0.22\textwidth]{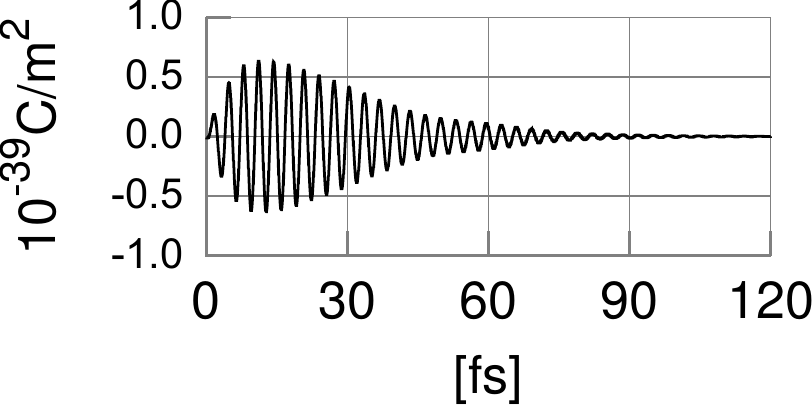}
        }
    \end{center}
    \caption{Transient response (polarization amplitude) of various monomers in the four-layer roll in response to initial $\hat{z}$ oriented electrical field of amplitude 1 V/m, imposed on monomers 1049 to 1108 located on a circular ring at the center of the innermost roll. The alphabetical labels in the legend correspond to the probe monomers depicted in \autoref{probes}.}
   \label{singleTrans}
\end{figure}

Considering the symmetric geometry of the roll and the initial excitation similar time signatures are expected to occur at both (top and bottom) ends of the roll. However, 
comparing `d' and `e' in \autoref{singleTrans} it appears that the polarization dynamics at the two ends of the chlorosome roll is slightly different. This can be linked to the chiral nature of the roll 
although a more systematic examination of the problem is needed before definitive conclusions can be made. Also as can be seen in \autoref{singleTrans}, the responses on the outmost sub-roll, i.e. sub-roll D, 
seem to have longer lifetimes which is likely to be consequence of larger radius of sub-roll D compared to sub-roll A.

The estimated number of molecules in the whole chlorosome is of the order of $200,000-250,000$~\cite{Oostergetel}, which is about 20 times larger that our four-layered roll model. 
In order to examine the performance of the presented parametric \hm technique for the structures comparable with the size of the chlorosome, we obtain the 
transient response of a 2 by 4 array of rolls with a lattice spacing of $25.5 \textnormal{nm}$ along both $\hat{x}$ and $\hat{y}$ directions. The resulting structure consists 
of $103680$ monomers and thus each frequency solution involves construction and $\mathcal{H}-\textnormal{compression}$  of a $103680\times103680$ matrix. Similar to \autoref{TSingle}, a central 
ring of monomers located in roll 1 of the array is exposed to initial electrical field along $\hat{z}$ direction and then the response of the system (polarization) at various monomers is 
presented in \autoref{4x2-1} and \autoref{4x2-2}.

\begin{figure}[TH]
     \begin{center}
        \subfigure[side view]
        {
            \includegraphics[width=0.35\textwidth]{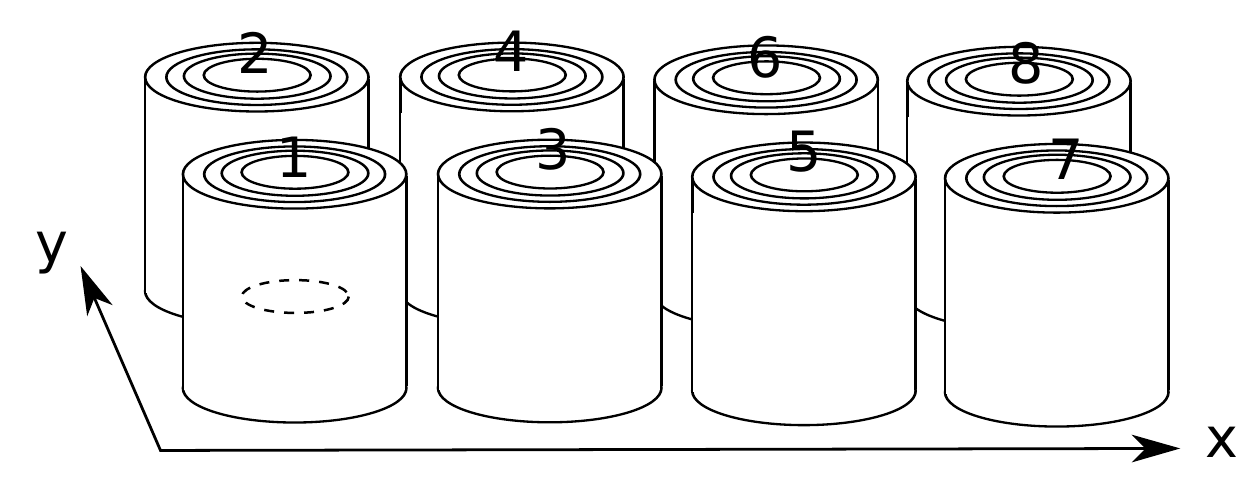}
        } \\
        \subfigure[top view]
        {
           \includegraphics[width=0.35\textwidth]{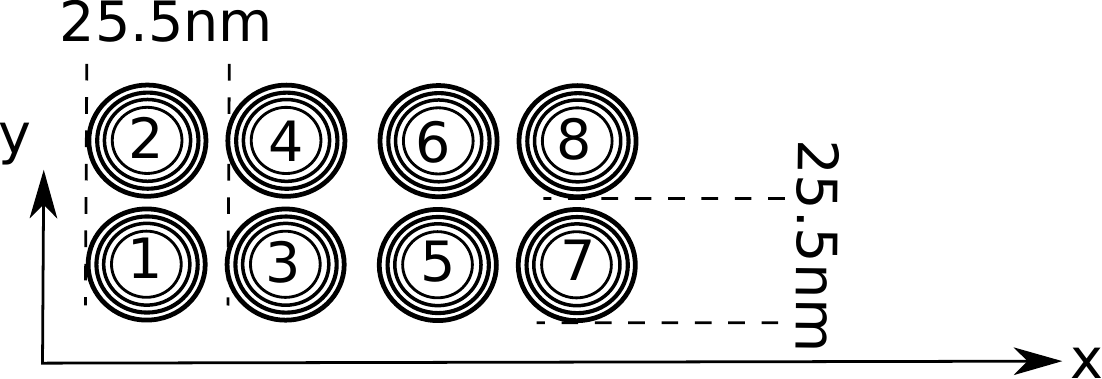}
        }
    \end{center}
    \caption{The 3D configuration of the $4\times2$ roll array. The dotted line shows the location of the monomers exposed to initial electrical field.}
   \label{arrayConf}
\end{figure}

\begin{figure}[TH]
     \begin{center}
        \subfigure[@ a]
        {
            \includegraphics[width=0.22\textwidth]{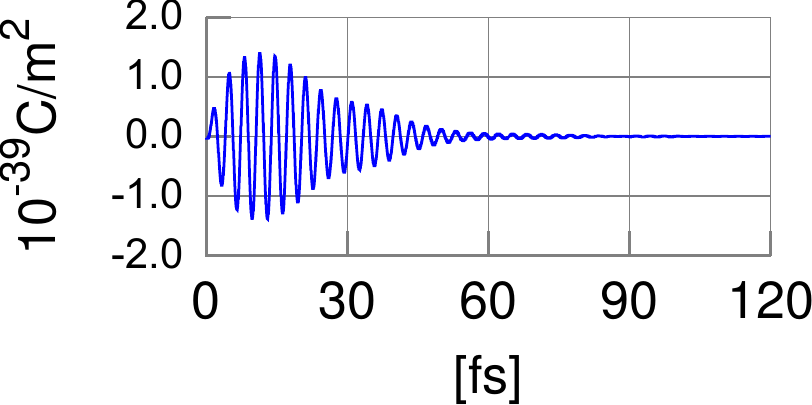}
        } \hfill
        \subfigure[@ d]
        {
           \includegraphics[width=0.22\textwidth]{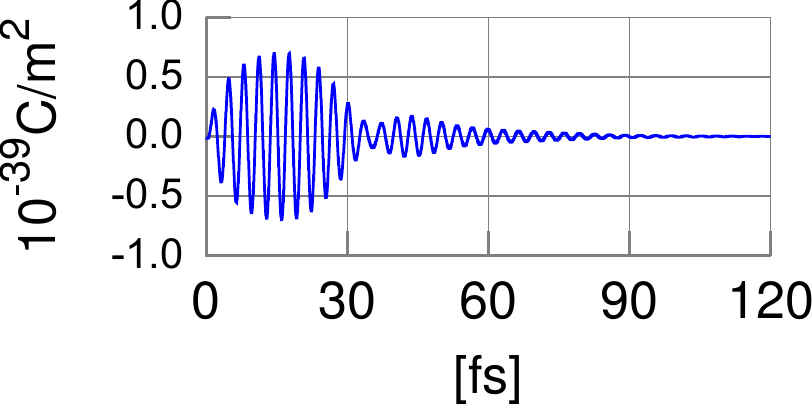}
        } \\
        \subfigure[@ b]
        {
           \includegraphics[width=0.22\textwidth]{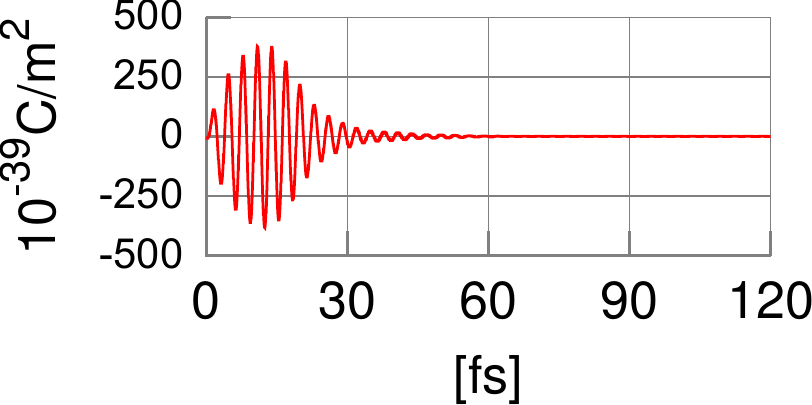}
        } \hfill
        \subfigure[@ e]
        {
           \includegraphics[width=0.22\textwidth]{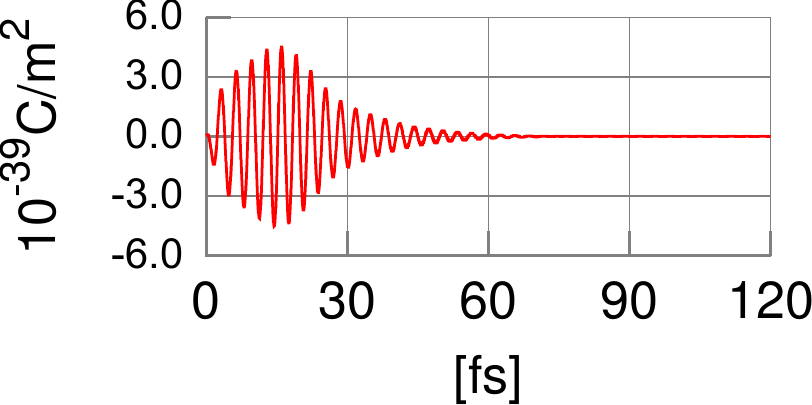}
        } \\
        \subfigure[@ c]
        {
           \includegraphics[width=0.22\textwidth]{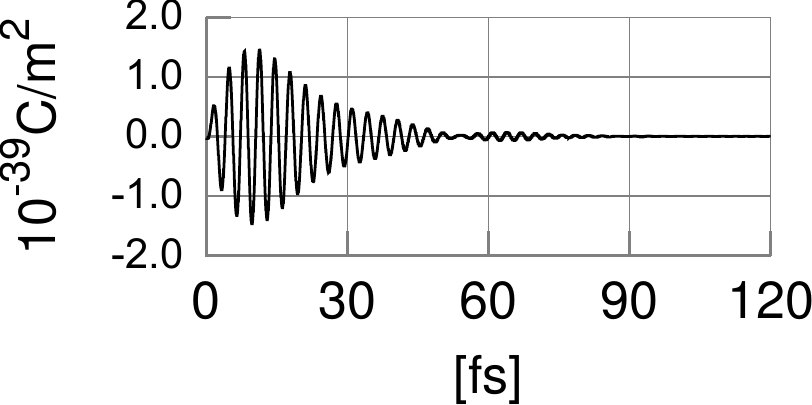}
        } \hfill
        \subfigure[@ f]
        {
           \includegraphics[width=0.22\textwidth]{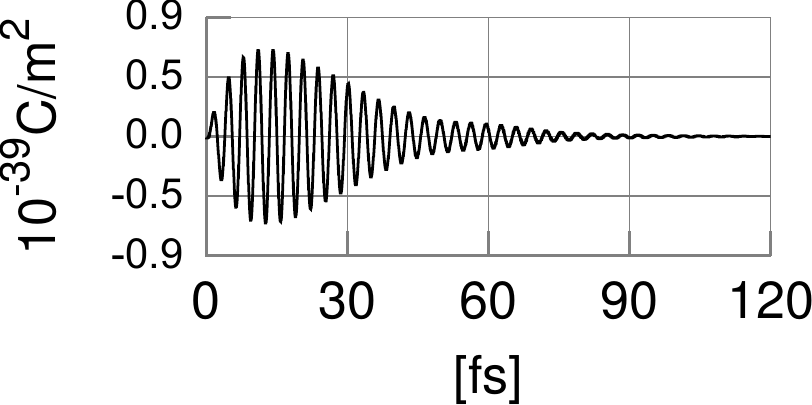}
        }
    \end{center}
    \caption{Transient response (polarization amplitude) of various monomers in chlorosome roll number $1$ in the array configuration illustrated in \autoref{arrayConf}. The labels in the legend refer to observation monomers depicted in \autoref{probes}. The response is due to initial $\hat{z}$ oriented electrical field of amplitude 1 V/m, imposed on monomers 1049 to 1108 located in chlorosome roll number $1$.}
   \label{4x2-1}
\end{figure}

\begin{figure}[TH]
     \begin{center}
        \subfigure[@ a']
        {
            \includegraphics[width=0.22\textwidth]{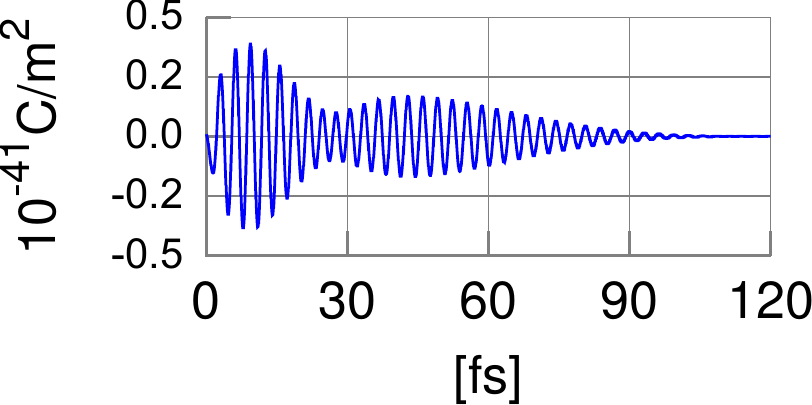}
        } \hfill
        \subfigure[@ d']
        {
           \includegraphics[width=0.22\textwidth]{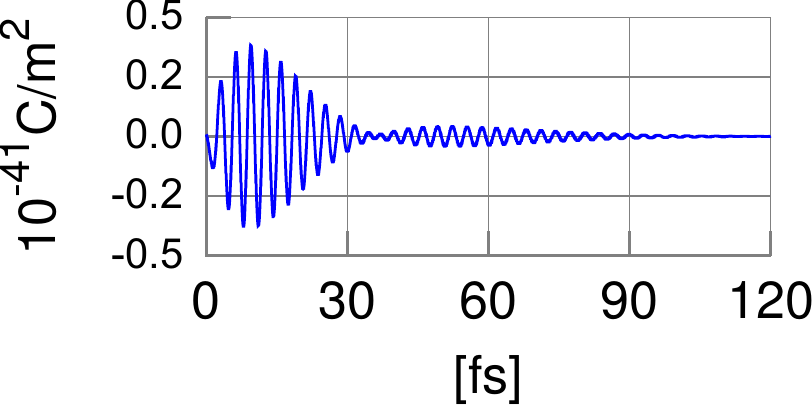}
        } \\
        \subfigure[@ b']
        {
           \includegraphics[width=0.22\textwidth]{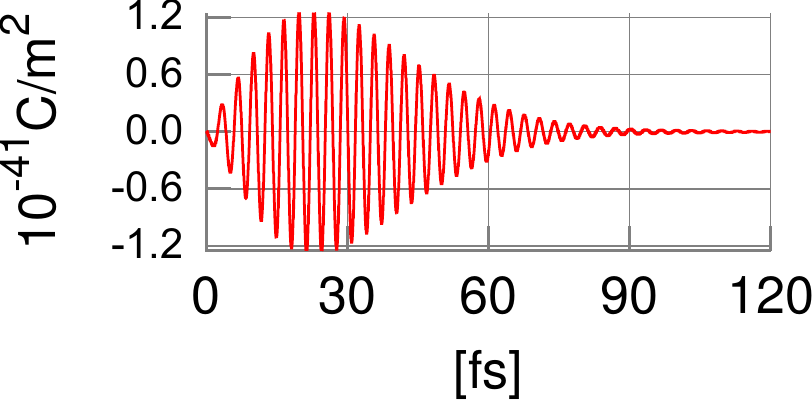}
        } \hfill
        \subfigure[@ e']
        {
           \includegraphics[width=0.22\textwidth]{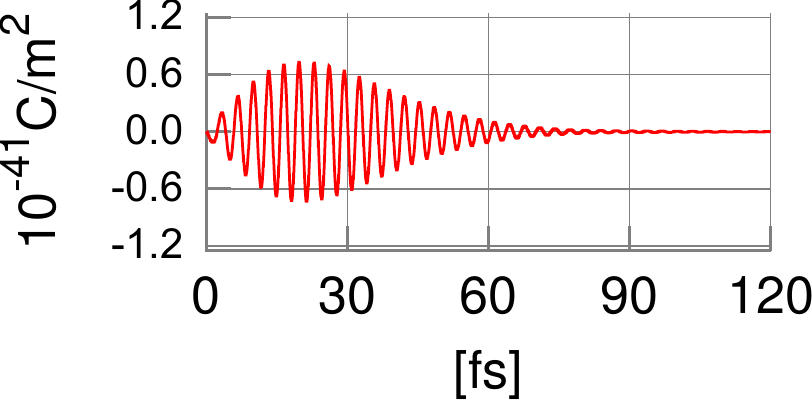}
        } \\
        \subfigure[@ c']
        {
           \includegraphics[width=0.22\textwidth]{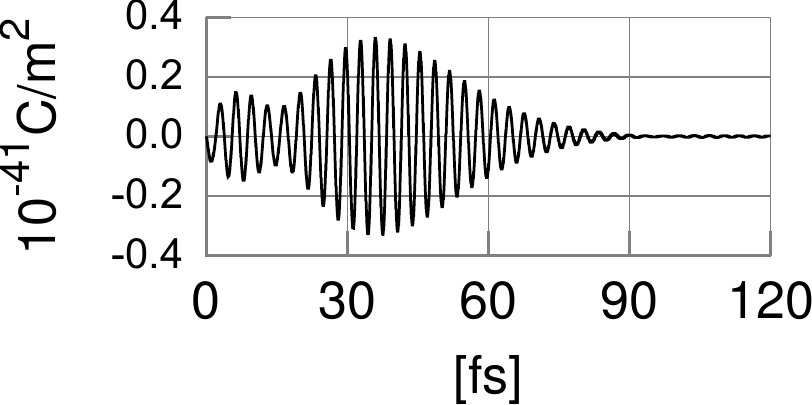}
        } \hfill
        \subfigure[@ f']
        {
           \includegraphics[width=0.22\textwidth]{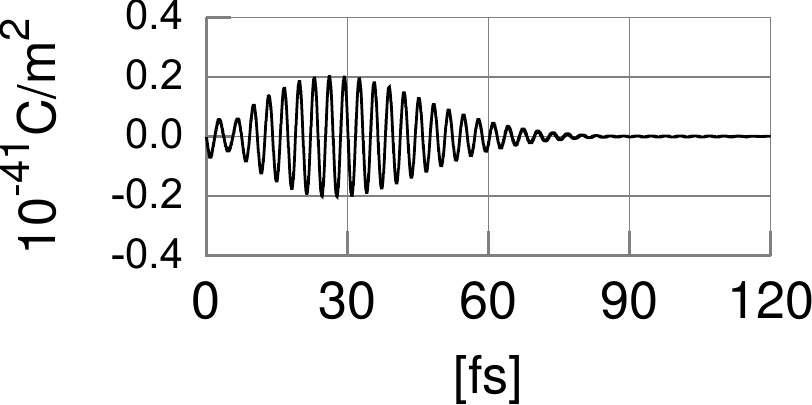}
        }
    \end{center}
    \caption{Transient response (polarization amplitude) of various monomers in chlorosome roll number $8$ in the array configuration illustrated in \autoref{arrayConf}. The labels in the legend refer to the observation monomers depicted in \autoref{probes} shifted into roll number $8$. The response is due to initial $\hat{z}$ oriented electrical field of amplitude 1 V/m, imposed on monomers 1049 to 1108 located in chlorosome roll number $1$.}
   \label{4x2-2}
\end{figure}

\subsubsection{Computational Statistics}

\begin{figure}[TH]
\includegraphics[width=0.475\textwidth]{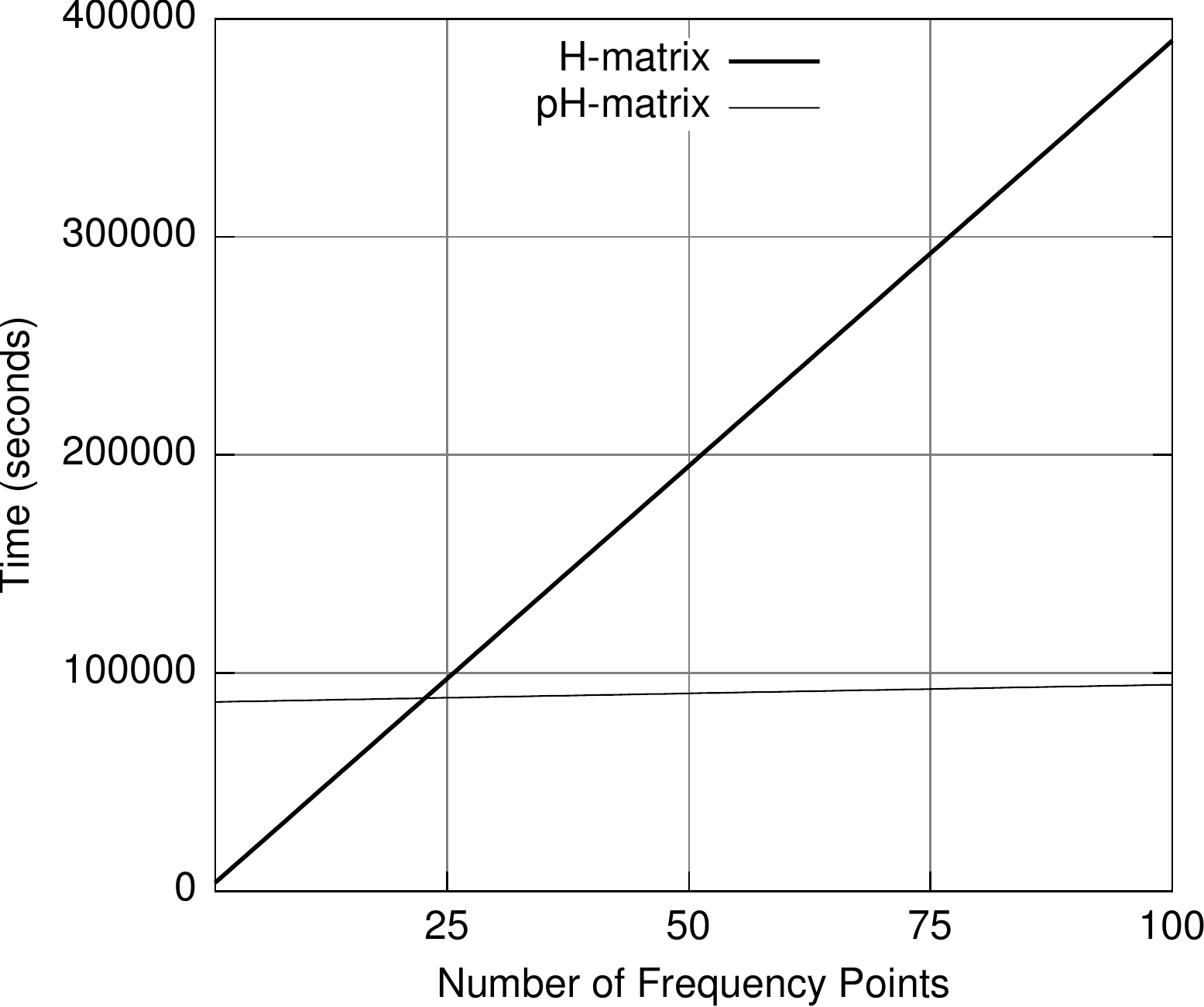}%
\caption{Comparison of the system matrix preparation times between the parametric and non-parametric \hm methods.}
\label{complexityComp}
\end{figure}

The statistics presented here correspond to the $2$ by $4$ array problem of \autoref{TSingle}.
A dense matrix representation in this case will require $103680\times103680\times16$ bytes (approximately 160 GB) of memory and thus
immediately ruled out. In a general scenario, the computational times can divided into two parts, (1) construction of the system matrix and (2) the iterative
solution of the problem. The iterative solution time is equal to the time required for one system matrix-vector multiply operation
times the number of iterations needed for the desired accuracy. Thus, the two methods have equal performance in this part.
An \hm representation of the system matrix using an accuracy threshold of $\epsilon = 10^{-4}$
leads to a compressed representation of the system that requires abtout $8GB$ of memory. 
However, the construction of the compressed \hm takes $65$ to $70$ minutes per frequency. 
Hence, the construction of the system \hm at $350$ frequency points requires $350\times 65 \times 60 = 1365000$ seconds.
On the other hand, using the parameterized \hm method, $13$ \hmp are constructed
at 13 equally spaced frequencies ranging from 100 THz to 700 THz and then the parametric \hm
is constructed from the samples. This process takes $13\times65\times 60$ seconds for the initial \hmp plus another $10$ hours for
the construction of the parametric \hm. Thus the parametric \hm requires a total of $86700$ seconds.
On top of that, at each frequency point, the \hm can be constructed from the parametric \hm in $80$ seconds leading to a speed up factor of $12$.
\Autoref{complexityComp} compares the growth of the computational times associated with the construction of the 
system matrix using both \hm and parametric \hm method. The curves depicted in \autoref{complexityComp}
are the representation of two linear equations with a considerable difference between their slopes and constant terms.
From the figure it can be seen that the parametric \hm method outperforms the \hm method when the problem 
needs to be solved at more than $20$ frequencies.

\section{Conclusion}\label{Conc}

An \hm acceleration method is adopted for the classical formulation arising in the simulation of molecular aggregates.
The \hm approach reduces the otherwise $\mathcal{O}(N^2)$ complexity that arises from direct implementation of the resulting matrix problem.
Moreover, a novel parametric \hm approach is introduced and provides an efficient means for the solution of large excitonic problems such as those encountered in the study 
of the photosynthesis process. A tubular aggregate of pigment molecules as used as an example to demonstrate that the developed method can 
give an order of magnitude acceleration in the calculations of transient responses as compared to the \hm approach. Further work on a detailed characterization of the physics of this model is in progress in our groups. Numerical experiments conducted in this work verify the validity of 
robustness of the method. Another important advantage of the parametric \hm method lies in the fact that it can easily be used for other kernels and thus other formulations including the 
quantum \emph{Hamiltonian} method can also be accelerated using this technique. The scope of the applicability of the parametric \hm method goes beyond the current application as it can be 
applied to many other \hm compatible problems including those arising from the discretization of integral operators in other fields such as acoustics, electromagnetics, fluid mechanics and 
fracture mechanics. Future works will focus on enhancing the method via computer parallelization. In this work, the parameterization was used for efficient frequency sweeping, however, 
the approach can potentially be applied to other types of parameter sweeping.

\section{Acknowledgements}
D. A.-O.-B., M. R.,E. C. and H. M. acknowledge US Air Force Office of Scientific Research (AFOSR) grant No. FA9550-10-1-0438 and in-part U.S. Office
of Naval Research (ONR) MURI award grant N00014-10-1-0942. S. K. S., S. V., and A. A.-G. acknowledge Defense Threat Reduction Agency grant HDTRA1-10-1-0046 and 
Defense Advanced Research Projects Agency grant N66001-10-1-4063. Further, A. A.-G. is grateful for the support from the Corning Foundation.

\bibliography{mybibs}

\end{document}